\documentclass[reqno,11pt]{amsart} 
\usepackage{amsmath}
\usepackage{amssymb}
\usepackage{amsthm}

\newcommand\NoBlackBoxes{\global\overfullrule0pt}
\NoBlackBoxes
\theoremstyle{plain}

\begin{document}

\def\theequation{\thesection.\arabic{equation}}
\def\E{{\bf E}}
\def\R{{\bf R}}
\def\C{{\bf C}}
\def\P{{\bf P}}
\def\Z{{\bf Z}}
\def\H{{\cal H}}
\def\Im{{\rm Im}}
\def\Tr{{\rm Tr}}

\def\k{{\kappa}}
\def\M{{\cal M}}
\def\Var{{\rm Var}}
\def\Ent{{\rm Ent}}
\def\O{{\rm Osc}_\mu}

\def\ep{\varepsilon}
\def\phi{\varphi}
\def\F{{\cal F}}
\def\L{{\cal L}}

\def\be{\begin{equation}}
\def\en{\end{equation}}
\def\bee{\begin{eqnarray*}}
\def\ene{\end{eqnarray*}}

\title{Fisher-type information involving \\
higher order derivatives}

\author{Sergey G. Bobkov$^{1}$}
\thanks{1) 
School of Mathematics, University of Minnesota, Minneapolis, MN, USA
}


\begin{abstract}
Basic general properties are considered for the Fisher-type information 
involving higher order derivatives. They are used to explore various
properties of probability densities and to derive Stam-type inequalities.
\end{abstract} 
 
\maketitle
\markboth{Sergey G. Bobkov}{Fisher-type information}

\def\theequation{\thesection.\arabic{equation}}
\def\E{{\mathbb E}}
\def\R{{\mathbb R}}
\def\C{{\mathbb C}}
\def\P{{\mathbb P}}
\def\Z{{\mathbb Z}}
\def\S{{\mathbb S}}
\def\I{{\mathbb I}}
\def\T{{\mathbb T}}

\def\s{{\mathbb s}}

\def\G{\Gamma}

\def\Ent{{\rm Ent}}
\def\var{{\rm Var}}
\def\Var{{\rm Var}}
\def\cov{{\rm cov}}
\def\V{{\rm V}}

\def\H{{\rm H}}
\def\Im{{\rm Im}}
\def\Tr{{\rm Tr}}
\def\s{{\mathfrak s}}
\def\A{{\mathfrak A}}
\def\m{{\mathfrak m}}

\def\k{{\kappa}}
\def\M{{\cal M}}
\def\Var{{\rm Var}}
\def\Ent{{\rm Ent}}
\def\O{{\rm Osc}_\mu}

\def\ep{\varepsilon}
\def\phi{\varphi}
\def\vp{\varphi}
\def\F{{\cal F}}

\def\be{\begin{equation}}
\def\en{\end{equation}}
\def\bee{\begin{eqnarray*}}
\def\ene{\end{eqnarray*}}

\thispagestyle{empty}

\maketitle


\section{{\bf Introduction}}
\setcounter{equation}{0}

\vskip2mm
\noindent
Given a random variable $X$ with an absolutely continuous density $f$, the
Fisher information hidden in the distribution of $X$ is defined by
\be
I(X) = \E\,\rho(X)^2 = \int_{-\infty}^{\infty} \frac{f'(x)^2}{f(x)}\,dx,
\en
where the integration is restricted to the set of points where $f(x)>0$. 
Here, $\rho = f'/f$ represents the logarithmic derivative of $f$, which is also 
called the score function (often being taken with the minus sign). Since 
$f(X)>0$ almost surely, the random variable $\rho(X)$, called the score 
of $X$, is well-defined and finite with probability one. 

The functional (1.1) has two natural generalizations motivated by
various problems in different fields. In particular, one is interested in the
behaviour of absolute moments of the scores
\be
I_p(X) = \E\,|\rho(X)|^p = 
\int_{-\infty}^{\infty} \frac{|f'(x)|^p}{f(x)^{p-1}}\,dx, \quad p \geq 1.
\en
As partial cases, the first absolute moment $I_1(X) = \|f\|_{\rm TV}$ 
describes the total variation norm of the density function $f$, and the second
moment is the Fisher information $I_2(X) = I(X)$.

Another closely related functional defined for positive integers $p$
is
\be
I^{(p)}(X) = \E\,\rho_p(X)^2 = \int_{f(x)>0} \frac{f^{(p)}(x)^2}{f(x)}\,dx.
\en
Here $\rho_p = f^{(p)}/f$ may be viewed as the ``$p$-th order" score function. 

These functionals were introduced 
by Lions and Toscani \cite{L-T} in their study of convergence of densities 
(and of their powers) in Sobolev spaces towards the central limit theorem. 
Previously, the functional $I_4$ was also considered by Gabetta \cite{G} 
in the context of the kinetic theory of gases to study the convergence 
to equilibrium in Kac's model. In the paper \cite{B1}, the moments of the scores
together with
exponential and Gaussian moments of $\rho(X)$ appear with the aim to control 
the translates of product probability measures. See also \cite{B2} and \cite{B3} 
for various upper bounds on the Fisher information and moments of the scores.

The quantity $I^{(p)}(X)$ may be called the Fisher(-type) information of order 
$p$. Denote by $\mathfrak C^p$ the collection of all continuous functions $f$
on the real line which have continuous derivatives up to order $p-1$, such that
$f^{(p-1)}$ is (locally) absolutely continuous. We denote by $f^{(p)}$
the derivative of $f^{(p-1)}$ which is defined and finite almost everywhere.

\vskip5mm
{\bf Definition 1.1.} If the random variable $X$ has a density $f$ 
from the class $\mathfrak C^p$ for an integer $p \geq 1$, the Fisher information 
$I^{(p)}(X) = I^{(p)}(f)$ of order $p$
is defined by (1.3). In all other cases, put $I^{(p)}(X) = \infty$.

\vskip5mm
Since $f^{(0)} = f$, it is natural to put $I^{(0)}(X) = 1$.

In this paper we explore general properties of the functional $I^{(p)}(X)$ 
and its relationship to various properties of densities $f$. Many of them 
extend and sharpen corresponding properties obtained under the hypothesis 
that the classical Fisher information $I(X)$ is finite. These properties include 
the integrability of the first $p$ derivatives of $f$ and assertions about their 
decay at infinity under moment assumptions posed on $X$. This will allow us
to consider the relative Fisher-type information with respect to the
standard normal distribution and to prove, for example, the following
comparison. In the sequel, we use the notation $Z \sim N(a,\sigma^2)$ 
for the case where
the random variable $Z$  is normal with mean $a$ and variance $\sigma^2$.

\vskip5mm
{\bf Theorem 1.2.} {\sl Let $I^{(p)}(X)$ be finite for an integer $p \geq 1$. Then,
for $Z \sim N(0,1)$,
\be
\E\,H_p(X)^2 = \E\,H_p(Z)^2 \ \Rightarrow \ I^{(p)}(X) \geq I^{(p)}(Z).
\en
}

Here and below $H_p$ denotes the Chebyshev-Hermite polynomial of 
degree $p$ with a leading coefficient 1
(let us note that the moment $\E X^{2p}$ should be finite as well). 
In the case $p=1$, (1.4) recovers a well-known statement that the Fisher 
information $I(X)$ is minimized for the normal distribution
when the variance is fixed. In other words, (1.4) may be viewed as a
generalization of the Cram\'er-Rao inequality for $I^{(p)}$.

One interesting question which we partly address is: How can one compare 
$I^{(p)}(X)$ for different $p$? For example, in the case of moments of 
the scores defined as in (1.2), the $L^p$-norms $p \rightarrow I_p(X)^{1/p}$ 
are non-decreasing. However, it may occur that the Fisher-type information 
is finite for a given odd order $p \geq 3$, while $I^{(q)}(X)$ are infinite for 
all even $q<p$ (cf. Example 2.5 below). Nevertheless, using the so-called 
isoperimetric profiles, one can derive the following relations for the case $p=2$.

\vskip5mm
{\bf Theorem 1.3.} {\sl For any random variable $X$,
\be
I^{(2)}(X) \geq \frac{1}{3}\,I_4(X) \geq \frac{1}{3}\,I(X)^2.
\en
}

Thus, the finiteness of $I^{(2)}(X)$ guarantees the finiteness of the usual
Fisher information.

Part of the proof of Theorem 1.3 is based on the lower semi-continuity 
of the Fisher-type information with respect to the weak convergence, 
as well as on the convexity of this functional
in the space of all probability distributions on the real line. These two important
properties reduce many relations such as (1.5) to the case
where $X$ has a $C^\infty$-smooth positive density on the real line, by means of the
following continuity property.

\vskip5mm
{\bf Theorem 1.4.} {\sl For all independent random variables $X$ and $Z$,
\be
\lim_{\ep \rightarrow 0} \, I^{(p)}(X + \ep Z) = I^{(p)}(X).
\en
}

In particular, if the distribution of $X$ is not absolutely continuous, then
$I^{(p)}(X + \ep Z) \rightarrow \infty$ regardless of whether 
or not $Z$ has a smooth density.

If $Z \sim N(0,1)$, then $I^{(p)}(X + \ep Z)$ is finite for any $\ep > 0$, and
the convergence in (1.6) is monotone in $\ep$. Hence, this equality may
be taken as an equivalent definition of $I^{(p)}(X)$, which was actually
proposed in \cite{L-T}.

The property (1.6) can be also used to study in full generality
generalizations of the classical Stam inequality (\cite{S}, \cite{D-C-T}, \cite{J})
\be
\frac{1}{I(X+Y)} \geq \frac{1}{I(X)} + \frac{1}{I(Y)}.
\en
In particular, we have:

\vskip5mm
{\bf Theorem 1.5.} {\sl Given independent random variables $X$ and $Y$,
for all $k = 1,\dots,p-1$, $p \geq 2$,
\be
\frac{1}{I^{(p)}(X+Y)} \geq \frac{1}{I^{(p)}(X)} + \frac{1}{I^{(p)}(Y)} + 
\frac{1}{I^{(k)}(X) I^{(p-k)}(Y)}.
\en
}

\vskip2mm
In the case $p=2$, the family (1.8) contains only one inequality, in which
an equality is attained for the class of normal distributions similarly to (1.7).

Thus, (1.7) is satisfied for all $I^{(p)}$ in place of $I$. Another immediate
consequence of (1.8) is that the finiteness of $I^{(k)}(X)$ and $I^{(p-k)}(Y)$ 
with $1 \leq k\leq p-1$ guarantees the finiteness of $I^{(p)}(X+Y)$ in view of
the following immediate consequence from (1.8)
$$
I^{(p)}(X+Y) \leq I^{(k)}(X) I^{(p-k)}(Y).
$$

By induction, it also follows that
$$
I^{(p)}(X_1 + \dots + X_p) \leq I(X_1) \dots I(X_p)
$$
whenever the  random variables $X_1,\dots,X_p$ are independent.
In this connection, let us recall that the convolution of 3 probability
densities with a finite total variation norm has a finite Fisher information
(\cite{B3, B-C-G}). Hence, the sum of $3p$ independent
random variables whose densities are functions of bounded total variation
has a finite Fisher-type information of order $p$.

In the proof of (1.8), we recall the argument by Lions and Toscani 
\cite{L-T}. However, in Lemma 2.3 they state a Stam-type 
inequality for the functional $I^{(p)}$ as a different relation 
$$
I^{(p)}(X+Y) \leq
\sum_{k=0}^p \alpha_k^2\, I^{(k)}(X) I^{(p-k)}(Y)
$$
with arbitrary $\alpha_i \geq 0$ such that $\alpha_0 + \dots + \alpha_p = 1$.
Optimizing over the coefficients $\alpha_i$, it is equivalent to
\be
\frac{1}{I^{(p)}(X+Y)} \, \geq \, \sum_{k=0}^p
\frac{1}{I^{(k)}(X) I^{(p-k)}(Y)},
\en
which is sharper than (1.8) for $p  \geq 3$ in view of the additional terms
on the right-hand side of (1.9). But, in order to reach (1.9),
it has to be required in the last step of the proof that
\be
\int_{-\infty}^\infty \frac{f^{(k)} f^{(l)}}{f}\,dx \, 
\int_{-\infty}^\infty \frac{g^{(p-k)} g^{(p-l)}}{g}\,dx \leq 0, 
\quad k \neq l \ \ (k,l = 1,\dots,p-1),
\en
for the densities $f$ and $g$ of $X$ and $Y$. Testing (1.10) in the class 
of the Gamma distributions with $p=3$, we have come to the conclusion 
that this is not correct. 

Nevertheless, there is a good reason to believe that the relation (1.9)
is true, as it becomes an equality for normal distributions with arbitrary
variances. Let us emphasize one particular case in this direction.

\vskip4mm
{\bf Theorem 1.6.} {\sl Let $X$ and $Y$ be independent random variables,
and let $X$ have a normal distribution. Then $(1.9)$ holds true.
}

\vskip4mm
Indeed, in the standard Gaussian case, the first integral in (1.10) is vanishing for
all $k \neq l$, which means the orthogonality of the Chebyshev-Hermite polynomials
in $L^2$ over the Gaussian measure. Hence the condition (1.10) is 
satisfied for any $g$.

We start with several examples illustrating the Fisher-type information
and then discuss basic properties of densities assuming that $I^{(p)}(X)$
is finite (Sections 2-5). A more general form of Theorem 1.2 is presented
in Section 6. Sections 7-10 contain detailed arguments towards the
lower semi-continuity, convexity and monotonicity of this functional, 
with proof of Theorem~1.4.
Sections 11-12 are aimed at proving Theorem 1.3, and the remaining Sections
13-15 deal with Stam-type inequalities. We use the following plan.

\vskip2mm
1. Introduction.

2. Examples.

3. First elementary properties.

4. Integrability of derivatives.

5. Polynomial decay of densities and their derivatives.

6. Relative Fisher information of order $p$.

7. Lower semi-continuity.

8. Convex mixtures of probability measures.

9. Convexity of the Fisher-type information.

10. Monotonicity and continuity along convolutions.

11. Representations in terms of isoperimetric profile.

12. Lower bounds for $I^{(2)}$ in terms of $I_4$ and $I$.

13. Stam-type inequality in the case $p \geq 2$.

14. Stam-type inequality with Gaussian components.

15. The Gamma distributions.


\vskip5mm
\section{{\bf Examples}}
\setcounter{equation}{0}

\vskip2mm
\noindent
It is useful to keep in mind that the functional 
$I^{(p)}$ is shift invariant and homogeneous of order $-2p$ 
with respect to $X$, that is,
$$
I^{(p)}(a+bX) = b^{-2p}\,I^{(p)}(X), \quad a \in \R, \ b \neq 0.
$$

\vskip2mm
{\bf Example 2.1.} If $Z \sim N(0,1)$, then $I(Z) = 1$. The standard normal density 
$$
f(x) = \varphi(x) = \frac{1}{\sqrt{2\pi}}\,e^{-x^2/2}
$$ 
of $Z$ has derivatives $f^{(p)}(x) = (-1)^p\,H_p(x) \varphi(x)$. 
Hence $\rho_p(x) = (-1)^p\,H_p(x)$ and
$$
I^{(p)}(Z) = \E\,H_p(Z)^2 = p!
$$
More generally, if $X \sim N(a,\sigma^2)$ with parameters 
$a \in \R$ and $\sigma>0$, then $I^{(p)}(X) = p! \, \sigma^{-2p}$.

Hence, the relation (1.9) specialized to independent random variables
$X \sim N(a_1,\sigma_1^2)$ and $Y \sim N(a_2,\sigma_2^2)$
becomes an equality in the binomial formula.

\vskip4mm
{\bf Example 2.2.} Let $X$ have a beta distribution with parameters 
$\alpha,\beta > 0$, i.e. with density
$$
f(x) = \frac{1}{B(\alpha,\beta)}\,x^{\alpha - 1} (1-x)^{\beta - 1}, \quad
0 < x <1.
$$
Near zero $\frac{f^{(p)}(x)^2}{f(x)} \sim {\rm const}\cdot x^{\alpha - 2p - 1}$ 
which is integrable in a neighborhood of zero, if and only if
$\alpha > 2p$. In this case, the derivatives are continuous at zero
for all $k = 0,1,\dots,p-1$. A similar conclusion is true about the point $x=1$,
and we conclude that
$$
I^{(p)}(X) < \infty \ \Longleftrightarrow \ \min(\alpha,\beta) > 2p.
$$

\vskip2mm
{\bf Example 2.3.} Suppose that the random variable $X$ has an even 
positive density $f$ on the real line, which is $C^\infty$-smooth and 
such that
$$
f(x) = cx^{-q}, \quad x \geq 1,
$$
with parameter $q>1$ for some constant $c>0$. In this case
$f^{(p)}(x) = c_1\, x^{-q-p}$ for $x \geq 1$, where $c_1 \neq 0$ does not
depend on $x$.
Hence $I^{(p)}(X) < \infty$ for all integers $p \geq 1$.

\vskip4mm
{\bf Example 2.4.} If $X$ has density $f(x) = x e^{-x^2/2}$ supported on
the half-axis $x>0$, then $f'(x)  = (1-x^2)\, e^{-x^2/2}$ and
$f''(x)  = (x^3-3x)\, e^{-x^2/2}$. Hence $I(X) = \infty$, while
$$
\int_0^\infty \frac{f''(x)^2}{f(x)}\,dx < \infty.
$$
Nevertheless, $I^{(2)}(X) = \infty$, since $f'$ is not continuous:
$f'(0-) = 0$, $f'(0+) = 1$.

\vskip4mm
{\bf Example 2.5.} Consider the $C^\infty$-smooth density
$$
f(x) = x^2 \varphi(x) = \varphi(x) + \varphi_2(x) = \frac{1}{\sqrt{2\pi}}\,
x^2 e^{-x^2/2}, \quad x \in \R,
$$
where we have involved the Hermite functions $\varphi_p(x) = H_p(x) \varphi(x)$. 
Using $\varphi_p' = -\varphi_{p+1}$, we have 
$f^{(p)} = (-1)^p\,(\varphi_p + \varphi_{p+2})$, so that
$$
\frac{f^{(p)}(x)^2}{f(x)} = \frac{(H_p(x) + H_{p+2}(x))^2}{x^2}\,\varphi(x).
$$
Whether or not this function is integrable is determined by the behavior of the last 
ratio near zero. Since 
$H_{2p}(0) + H_{2p+2}(0) = c_p$ and $H_{2p-1}(x) + H_{2p+1}(x) \sim -c_p x$
as $x \rightarrow 0$ with non-zero constants $c_p = (-1)^{p-1}\,\frac{(2p)!}{(p-1)!\, 2^{p-1}}$, 
we conclude that
$$
I^{(2p-1)}(X) < \infty, \ \ I^{(2p)}(X) = \infty \quad (p \geq 1).
$$

\vskip2mm
{\bf Example 2.6.} Let $X$ have a Gamma distribution with $n$ degrees of
freedom, that is, with density
$$
f(x) = \frac{x^{n-1}}{\Gamma(n)}\,e^{-x}, \quad x > 0
$$
(where $n$ may be a real positive number).
Similarly to the beta distributions, $I^{(p)}(X)$ is finite if and only if $n>2p$.
For the first three values of $p$, direct computations show that
\begin{eqnarray}
I(X) 
 & = &
\frac{1}{n-2}, \\
I^{(2)}(X) 
 & = &
\frac{2(n+2)}{(n-2)(n-3)(n-4)}, \\
I^{(3)}(X) 
 & = &
\frac{6\,(n^2 + 13 n + 6)}{(n-2)(n-3)(n-4)(n-5)(n-6)} \nonumber
\end{eqnarray}
for the parameters $n>2$, $n>4$, and $n > 6$, respectively
(the formula (2.1) was already mentioned in \cite{J}).
We postpone the derivation of (2.1)-(2.2) to Section 15.

\vskip7mm
\section{{\bf First elementary properties}}
\setcounter{equation}{0}

\vskip2mm
\noindent
It is well-known that, if $I(X)$ is finite, then the density $f$ 
of $X$ represents a function of bounded variation on the real line 
with a total variation norm satisfying
$$
\|f\|_{\rm TV} = \int_{-\infty}^\infty |f'(x)|\,dx = 
\E\,|\rho(X)| \leq \sqrt{I(X)}.
$$
In particular, $f(-\infty) = f(\infty) = 0$, and $f$ is bounded by
$\sqrt{I(X)}$. The latter implies
$$
\int_{-\infty}^\infty f'(x)^2\,dx \leq I(X)^{3/2}.
$$

We now extend these relations to the Fisher-type information functionals of 
orders $p \geq 1$. Here and in the sequel, the following elementary observation
will be needed.

\vskip5mm
{\bf Proposition 3.1.} {\sl Let $I^{(p)}(X)$ be finite. If $f(x) = 0$ at
the point $x \in \R$ and $f^{(p-1)}$ has a finite derivative $f^{(p)}(x)$, 
then necessarily $f^{(p)}(x) = 0$. We also have $f'(x) = 0$.
}

\vskip5mm
{\bf Proof.} Since $f$ is non-negative, necessarily $f'(x) = 0$, and we are
done in the case $p=1$. If $p \geq 2$, then, by Taylor's formula 
in the Peano form,
$$
f(x+h) = \frac{a_2}{2!}\, h^2 + \dots + \frac{a_p}{p!}\, h^p + o(|h|^p), \qquad 
a_k = f^{(k)}(x), \  2 \leq k \leq p,
$$
and $f^{(p)}(x) = a_p + o(|h|)$ as $h \rightarrow 0$. Assuming that
$f^{(p)}(x) \neq 0$, let $k$ be the smallest integer in the interval
$2 \leq k \leq p$ such that $f^{(k)}(x) \neq 0$. Then $a_k \neq 0$, 
$a_p \neq 0$, so that
$$
\frac{f^{(p)}(x+h)^2}{f(x+h)} = 
\frac{a_p^2 + o(|h|)}{\frac{a_k}{k!}\,h^k + o(|h|^k)}.
$$
But this function is not integrable over $h \in (-\ep,\ep)$ with $\ep>0$ small enough.
\qed

\vskip5mm
{\bf Proposition 3.2.} {\sl If $I^{(p)}(X)$ is finite, the derivative
$f^{(p-1)}$ represents a function of bounded variation
with a total variation norm
\be
\|f^{(p-1)}\|_{\rm TV} = \int_{-\infty}^\infty |f^{(p)}(x)|\,dx 
\leq \sqrt{I_p(X)}.
\en
In particular, $f^{(p-1)}(-\infty) = f^{(p-1)}(\infty) = 0$, and 
$$
\max_x |f^{(p-1)}(x)| \leq \sqrt{I^{(p)}(X)}.
$$
}

{\bf Proof.} By the assumption, the derivative $f^{(p-1)}$ is a locally 
absolutely continuous function. Hence, it is differentiable 
on a set $E \subset \R$ of full Lebesgue measure, and we have
an equality in (3.1) for its derivative $f^{(p)}$. By Proposition 3.1, 
$f^{(p)}(x) \neq 0 \Rightarrow f(x) > 0$ for all
$x \in E$. Applying the Cauchy inequality, we obtain that
\bee
\int_{-\infty}^\infty |f^{(p)}(x)|\,dx 
 & = &
\int_{f(x) > 0} |f^{(p)}(x)|\,dx \\
 & = &
\int_{f(x) > 0} \frac{|f^{(p)}(x)|}{\sqrt{f(x)}}\,\sqrt{f(x)}\,dx \ \leq \
\sqrt{I^{(p)}(X)},
\ene
proving the first assertion. As a consequence, the limits
$$
f^{(p-1)}(-\infty) = \lim_{x \rightarrow -\infty} f^{(p-1)}(x), \qquad
f^{(p-1)}(\infty) = \lim_{x \rightarrow \infty} f^{(p-1)}(x)
$$
exist and are finite. Necessarily, these limits must be zero, since otherwise
$f(x)$ would behave polynomially at infinity contradicting to the integrability
of $f$. Finally,
$$
\max_x |f^{(p-1)}(x)| \leq \|f^{(p-1)}\|_{\rm TV} \leq \sqrt{I^{(p)}(X)}.
$$
\qed

\vskip2mm
{\bf Proposition 3.3.} {\sl If $I^{(p)}(X)$ is finite, then
$$
\int_{-\infty}^\infty |f^{(p)}(x)|^2\,dx \leq I^{(p)}(X)^{3/2}.
$$
}

This follows from
$$
\int_{-\infty}^\infty |f^{(p)}(x)|^2\,dx \leq 
\max_x f(x) \int_{f(x) > 0} \frac{|f^{(p)}(x)|^2}{f(x)}\,dx.
$$

\vskip5mm
\section{{\bf Integrability of derivatives}}
\setcounter{equation}{0}

\vskip2mm
\noindent
Applying Proposition 3.2, one may extend its bound on the total variation
norm to all derivatives smaller than $p$ (in a certain form). 
As before, we assume that $p \geq 1$ is an integer.

\vskip5mm
{\bf Proposition 4.1.} {\sl If $f$ is the density of a random variable $X$
with finite $I^{(p)}(X)$, then all derivatives $f^{(k)}$, $1 \leq k \leq p$, 
are integrable functions. Moreover,
\be
\|f^{(k-1)}\|_{\rm TV} \, = \, \int_{-\infty}^\infty |f^{(k)}(x)|\,dx 
 \, \leq \, C_p\, I^{(p)}(X)^{\frac{k}{2p}}
\en
with some constants $C_p$ depending on $p$ only.
In particular, if $f$ is supported on the interval $(a,b)$, finite or not, then
$f^{(k-1)}(a+) = f^{(k-1)}(b-) = 0$. In addition,
\be
\max_x |f^{(k-1)}(x)| \leq C_p\, I^{(p)}(X)^{\frac{k}{2p}}.
\en
}

Before turning to the proof, let us mention two immediate consequences.

\vskip5mm
{\bf Corollary 4.2.} $I^{(p)}(X) > 0$.

\vskip4mm
Indeed, in the case $I^{(p)}(X) = 0$, we would obtain from (4.1) with 
$k=1$ that $\|f\|_{\rm TV} = 0$. But this is only possible when $f$ is a constant.

Another immediate consequence from Proposition 4.1 concerns the decay of
the characteristic function
$$
\widehat f(t) = \E\,e^{itX} = \int_{-\infty}^\infty e^{itx} f(x)\,dx, \quad
t \in \R.
$$

\vskip2mm
{\bf Corollary 4.3.} {\sl If $I^{(p)}(X)$ is finite, then
$\widehat f(t) = o(|t|^{-p})$ as $|t| \rightarrow \infty$.
}

\vskip5mm
For the proof, one may repeatedly integrate by parts with $t \neq 0$, which gives
$$
\widehat f(t) \, = \, - 
\frac{1}{it} \int_{-\infty}^\infty e^{itx} f'(x)\,dx \, = \, \dots \, = \,
\Big(-\frac{1}{it}\Big)^p \int_{-\infty}^\infty e^{itx} f^{(p)}(x)\,dx.
$$
Here we used the property that all derivatives $f^{(k)}$ up to 
order $p$ are integrable and vanishing at infinity for all $k \leq p-1$. 
Since $f^{(p)}$ is integrable, the last integral tends to zero as
$|t| \rightarrow \infty$, by the Riemann-Lebesgue lemma.

\vskip5mm
{\bf Lemma 4.4.} {\sl For any integrable function $u$ having derivatives up 
to order $p \geq 2$ (in the Radon-Nikodym sense for the $p$-th derivative),
for all integers $1 \leq k \leq p-1$, 
\be
\int_{-\infty}^\infty |u^{(k)}(x)|\,dx \leq 
A_p \int_{-\infty}^\infty |u(x)|\,dx + 
B_p \int_{-\infty}^\infty |u^{(p)}(x)|\,dx
\en
with coefficients $A_p$ and $B_p$ depending on $p$ only, for example, with
$A_p = 4^{p-1}$ and $B_p = 2^{4^p}$.
}

\vskip5mm
{\bf Proof.}
The integrability of the derivatives $u^{(k)}$ is stated without proof in \cite{B3}. 
Assuming that $u$ and $u^{(p)}$ are integrable functions, the inequality (4.3) 
can be obtained by the repeated application of its particular case $p=2$, namely
\be
\int |u'|\,dx \leq \int |u| + \frac{2}{3} \int |u''|,
\en
which is derived for the class $\mathfrak C^{(2)}$ in \cite{B3}, Proposition 5.1.
Here and below the integration is carried out over the real line with respect to
the Lebesgue measure.

Applying (4.4) to $u'$ in place of $u$ and then (4.4) once more, we obtain that
$$
\int |u''| \, \leq \, \int |u'| + \frac{2}{3} \int |u'''| \, \leq \,
\int |u| + \frac{2}{3} \int |u''| + \frac{2}{3} \int |u'''|,
$$
which is solved as
\be
\int |u''| \leq 3\int |u| + 2 \int |u'''|.
\en
Moreover, an application of (4.5) in (4.4) yields
\be
\int |u'| \leq 3\int |u| + \frac{4}{3} \int |u'''|.
\en
Thus, the relation (4.3) holds true with
$A_2 = 1$, $B_2 = \frac{2}{3}$, and $A_3 = 3$, $B_3 = 2$.

In order to extend such inequalities to derivatives of higher orders, one may 
argue by induction on $p$.
To make an induction step from $p-1$ to $p$ with $p \geq 4$, suppose that
\be
\int |u^{(k)}| \leq 
A_{p-1} \int |u| + B_{p-1} \int |u^{(p-1)}|, \quad 1 \leq k \leq p-2.
\en
By (4.4) applied to $u^{(p-2)}$ in place of $u$ 
and then to the derivative of the smaller order, 
\bee
\int |u^{(p-1)}|
 & \leq &
\int |u^{(p-2)}| + \frac{2}{3} \int |u^{(p)}| \\
 & \leq &
\int |u^{(p-3)}| + \frac{2}{3} \int |u^{(p-1)}| + \frac{2}{3} \int |u^{(p)}|,
\ene
which is solved similarly to (4.5) as
\be
\int |u^{(p-1)}| \leq 3 \int |u^{(p-3)}| + 2 \int |u^{(p)}|.
\en
In order to estimate the intermediate integral in (4.8), it is natural 
to apply the induction hypothesis (4.7) with $k=p-3$, that is, 
\be
\int |u^{(p-3)}| \leq A_{p-1} \int |u| + B_{p-1} \int |u^{(p-1)}|.
\en
One may use this in (4.8) in order to solve the resulting inequality for the integral 
containing the derivative $u^{(p-1)}$. However, this is only possible when 
the coefficient $3B_{p-1}$ in front of this integral is smaller than 1. Since this 
is not the case, we need to modify (4.8), by applying this relation
to the function $u(\lambda x)$ instead of $u(x)$ with parameter $\lambda>0$.
Then we get
$$
\int |u^{(p-1)}| \leq
\frac{3}{\lambda^2} \int |u^{(p-3)}| + 2 \lambda \int |u^{(p)}|.
$$
Using (4.9), we get
$$
\int |u^{(p-1)}| \, \leq \, \frac{3}{\lambda^2}\,
\bigg[A_{p-1} \int |u| + B_{p-1} \int |u^{(p-1)}|\bigg] +
2 \lambda \int |u^{(p)}|.
$$
Let us choose here $\lambda = 2 \sqrt{B_{p-1}}$ which leads to
$$
\int |u^{(p-1)}| \, \leq \, \frac{3 A_{p-1}}{4 B_{p-1}} \int |u| + 
\frac{3}{4} \int |u^{(p-1)}| + 4 \sqrt{B_{p-1}} \int |u^{(p)}|,
$$
implying that
\be
\int |u^{(p-1)}| \leq
\frac{3 A_{p-1}}{B_{p-1}} \int |u| + 16 \sqrt{B_{p-1}} \int |u^{(p)}|.
\en
This is of the desired form for $k=p-1$.

If $1 \leq k \leq p-2$, we involve the induction hypothesis (4.7), which 
together with (4.10) gives
\be
\int |u^{(k)}| \leq 4 A_{p-1} \int |u| + 16\, B_{p-1}^{3/2} \int |u^{(p)}|.
\en
If we require that $B_{p-1} \geq 3/4$ (which is the case in (4.5)-(4.6)) 
and compare the coefficients in front of $\int |u|$ in (4.10)-(4.11), 
one may choose $A_p = 4A_{p-1}$ and hence $A_p = 4^{p-1}$ fits.
We also obtain the recurrent equation
$$
B_p = 16\, B_{p-1}^{3/2}.
$$
Let us put $B_p = 2^{b_p}$ and rewrite this equation as
$b_p = 4 + \frac{3}{2}\,b_{p-1}$. By induction on $p$, it follows that
$b_p \leq 2^{2p}$.
\qed

\vskip5mm
{\bf Proof of Proposition 4.1.} The case $k=p$ is governed by 
Proposition 3.2, so, we may assume that $1 \leq k \leq p-1$ with 
$p \geq 2$. Applying (4.3) to $u(x) = f(\lambda x)$, $\lambda>0$, we get
$$
\int |f^{(k)}| \, \leq \, A_p\, \lambda^{-k} + B_p\, \lambda^{p-k}
\int |f^{(p)}|.
$$
The optimization over all $\lambda$ yields
$$
\int |f^{(k)}| \leq C_p\,\Big(\int |f^{(p)}|\Big)^{k/p}
$$
with $p$-dependent constants $C_p$. It remains to apply Proposition 3.2.
\qed

\vskip7mm
\section{{\bf Polynomial decay of densities and their derivatives}}
\setcounter{equation}{0}

\vskip2mm
\noindent
If the moment $\beta_{2s} = \E\,|X|^{2s}$ is finite for some 
real number $s>0$, then (\cite{B-C-G}, Proposition 7.1)
$$
\int_{-\infty}^\infty |x|^s\, |f'(x)|\,dx \leq \sqrt{\beta_{2s}\, I(X)}.
$$
Moreover,
$$
\lim_{|x| \rightarrow \infty} (1 + |x|^s)\,f(x) = 0.
$$
These results may be generalized, which allows one to control a polynomial 
decay of densities and their derivatives at infinity.

\vskip5mm
{\bf Proposition 5.1.} {\sl If $I^{(p)}(X)$ and $\beta_{2s}$ are finite for an
integer $p \geq 1$ and real $s>0$, then
$$
\int_{-\infty}^\infty |x|^s\, |f^{(p)}(x)|\,dx  \leq 
\sqrt{\beta_{2s}\, I^{(p)}(X)}.
$$
As a consequence, for all $x \in \R$,
$$
|f^{(p-1)}(x)| \leq \frac{c}{1 + |x|^s}, \quad 
c = \big(1 + \sqrt{\beta_{2s}}\big) \sqrt{I^{(p)}(X)}.
$$
Moreover,
$$
f^{(p-1)}(x) = o\big(|x|^{-s}\big) \ \ {\sl as} \ 
|x| \rightarrow \infty.
$$
}

{\bf Proof.} Put $I = I^{(p)}(X)$. Recall that, by Proposition 3.1,
$f^{(p)}(x) \neq 0 \Rightarrow f(x) > 0$ for all points $x$ from
a set of  full Lebesgue measure. Hence, applying the Cauchy inequality, 
we have
\bee
\int_{-\infty}^\infty |x|^s\,|f^{(p)}(x)|\,dx 
 & = &
\int_{f(x) > 0} |x|^s\,|f^{(p)}(x)|\,dx \\
 & = &
\int_{f(x) > 0} 
\frac{|f^{(p)}(x)|}{\sqrt{f(x)}}\,|x|^s \sqrt{f(x)}\,dx \ \leq \
\sqrt{\beta_{2s} I}.
\ene
This proves the first assertion. 

Let us combine the obtained inequality with the one of
Proposition 3.2. Then we get
$$
\int_{-\infty}^\infty (1 + |y|^s)\, |f^{(p)}(y)|\,dy \leq 
\big(1 + \sqrt{\beta_{2s}}\big)\, \sqrt{I}.
$$
Restricting the integration on the left-hand side to the half-axis 
$y \geq x \geq 0$, the left integral can be bounded from below by
$$
(1 + |x|^s)\, \ep(x), \quad {\rm where} \ \ 
\ep(x) = \int_x^\infty |f^{(p)}(y)|\,dy. 
$$
Hence, for any $b>x$,
\bee
|f^{(p-1)}(x) - f^{(p-1)}(b)| 
 & = &
\Big|\int_x^b f^{(p)}(y)\,dy\Big| \\
 & \leq &
\int_x^\infty |f^{(p)}(y)|\,dy \ \leq \ 
\frac{1}{1 + |x|^s}\,\big(1 + \sqrt{\beta_{2s}}\big) \sqrt{I}.
\ene
Letting $b \rightarrow \infty$ and applying the property 
$f^{(p-1)}(b) \rightarrow 0$ (Proposition 3.2), we arrive at 
the second required inequality. Since $\ep(x) \rightarrow 0$ as 
$x \rightarrow \infty$, the last assertion follows as well.
\qed

\vskip5mm
{\bf Proposition 5.2.} {\sl If $I^{(p)}(X)$ and $\beta_{2s}$ are finite 
for an integer $p \geq 1$ and $s > 0$, then
$$
f^{(p-k)}(x) = o\Big(\frac{1}{|x|^{s-k+1}}\Big), \quad 
k = 1,\dots,p, \ \ s > k-1,
$$
as $|x| \rightarrow \infty$. Moreover, in the case $k=p$,
$$
f(x) = o\Big(\frac{1}{|x|^{s-p+1}}\Big), \quad s \geq p-1.
$$
}

{\bf Proof.} The case $k = 1$ corresponds to Proposition 5.1:
$$
|f^{(p-1)}(y)| \leq \frac{\ep(y)}{1 + |y|^s},
$$
where $\ep(y) \rightarrow 0$ as $|y| \rightarrow \infty$. After the repeated
integration of this inequality over $y > x \geq 0$, and using
$f^{(p-l)}(\infty) = 0$, $1 \leq l \leq p$ (Proposition 4.1), we get, 
as $x \rightarrow \infty$,
\bee
f^{(p-2)}(x) 
 & = &
o\big(x^{-(s-1)}\big), \qquad p \geq 2, \ s > 1, \\ 
f^{(p-3)}(x)
 & = &
o\big(x^{-(s-2)}\big), \qquad p \geq 3, \ s > 2, \ \ldots \\
f^{(p-k)}(x)
 & = &
o\big(x^{-(s-(k-1))}\big), \quad p \geq k, \ s > k-1,
\ene
which corresponds to the first claim. In the remaining case $k = p$ and 
$s = p-1$, the second claim $f(x) = o(1)$ holds true 
according to Proposition 4.1.
\qed

\vskip7mm
\section{{\bf Relative Fisher information of order $p$}}
\setcounter{equation}{0}

\vskip2mm
\noindent
Given two random variables $X$ and $Y$ with densities $f$ and $g$
from the class $\mathfrak C^p$, define the relative Fisher 
information of an integer order $p \geq 1$ by
$$
I^{(p)}(X|Y) = I^{(p)}(f|g) = \int_{-\infty}^\infty 
\Big|\frac{f^{(p)}(x)}{f(x)} - \frac{g^{(p)}(x)}{g(x)}\Big|^2\,f(x)\,dx.
$$
This is a natural extension of the classical order $p=1$
(see also \cite{T} for other extensions).

Of a special interest is the case $Y=Z$ with the standard normal density
$g = \varphi$. Then
$$
I^{(p)}(X|Z) = I^{(p)}(f|\varphi) = \int_{-\infty}^\infty 
\Big|\frac{f^{(p)}(x)}{f(x)} - (-1)^p H_p(x)\Big|^2\,f(x)\,dx.
$$
Since the Chebyshev-Hermite polynomial $H_p(x)$ has degree $p$, 
for the last integral to be finite it is natural to require that $X$ have 
a finite moment $\beta_{2p}(X) = \E\,X^{2p}$. Then, opening the brackets, 
we get another representation
\be
I^{(p)}(X|Z) = I^{(p)}(X) - 2\, (-1)^p \int_{-\infty}^\infty 
f^{(p)}(x) H_p(x)\,dx + \E\,H_p(X)^2.
\en
Assuming that $I^{(p)}(X)$ is finite, the above integral is finite according
to Proposition 5.1 and may be easily evaluated. Namely, by Proposition 5.2 
with $s=p$,
$$
f^{(p-k)}(x) = o\Big(\frac{1}{|x|^{p-k+1}}\Big) \quad 
{\rm as} \ |x| \rightarrow \infty, \quad k = 1,\dots,p-1.
$$
Hence, integrating by parts and using $H_p'(x) = p H_{p-1}(x)$, we have
\bee
(-1)^p \int_{-\infty}^\infty  f^{(p)}(x) H_p(x)\,dx 
 & = &
(-1)^p \int_{-\infty}^\infty  H_p(x)\,df^{(p-1)}(x) \\
 & = &
(-1)^{p-1} \int_{-\infty}^\infty f^{(p-1)}(x)\, dH_p(x) \\
 & = &
(-1)^{p-1}\, p \int_{-\infty}^\infty f^{(p-1)}(x) H_{p-1}(x)\,dx. \\
\ene
Repeating the integration by parts, we will arrive at
\be
(-1)^p \int_{-\infty}^\infty  f^{(p)}(x) H_p(x)\,dx = p!
\int_{-\infty}^\infty f(x) H_0(x)\,dx = p!
\en
The latter factorial may also be written as $I^{(p)}(Z) = \E\,H_p(Z)^2$.

Applying (6.2) in (6.1), we arrive at the following assertion containing Theorem 1.2.

\vskip5mm
{\bf Proposition 6.1.} {\sl If $I^{(p)}(X)$ and $\beta_{2p}(X)$ are finite 
for an integer $p \geq 1$, then
$$
I^{(p)}(X|Z) = I^{(p)}(X) - 2 p! + \E\,H_p(X)^2.
$$
In particular,
$$
I^{(p)}(X) + \E\,H_p(X)^2 \geq 2 p!
$$
with equality if and only if $X$ has a standard normal distribution.
Therefore,
$$
\E\,H_p(X)^2 = \E\,H_p(Z)^2 \ \Rightarrow \ I^{(p)}(X) \geq I^{(p)}(Z).
$$
}

One may generalize this statement by replacing $H_p(x)$ with 
an arbitrary polynomial $H(x) = x^p + a_{p-1} x^{p-1} + \dots + a_0$ 
with leading coefficient 1. In this case, similarly to (6.1)
\bee
\int_{-\infty}^\infty 
\Big|\frac{f^{(p)}(x)}{f(x)} - (-1)^p\, H(x)\Big|^2\,f(x)\,dx
 & = &
I^{(p)}(X) \\
 & & \hskip-20mm
- 2\, (-1)^p \int_{-\infty}^\infty 
f^{(p)}(x) H(x)\,dx + \E\,H(X)^2,
\ene
while, integrating by parts, we have
\bee
(-1)^p \int_{-\infty}^\infty  f^{(p)}(x) H(x)\,dx 
 & = &
(-1)^p \int_{-\infty}^\infty  H(x)\,df^{(p-1)}(x) \\
 & = &
(-1)^{p-1} \int_{-\infty}^\infty f^{(p-1)}(x)\, dH(x) \\
 & = &
(-1)^{p-1} \int_{-\infty}^\infty f^{(p-1)}(x) H'(x)\,dx. \\
\ene
Repeating the integration by parts, we will arrive at
$$
(-1)^p \int_{-\infty}^\infty  f^{(p)}(x) H(x)\,dx = 
\int_{-\infty}^\infty  f(x) H^{(p)}(x)\,dx = p!
$$
Hence, we arrive at:

\vskip5mm
{\bf Proposition 6.2.} {\sl If $I^{(p)}(X)$ and $\beta_{2p}(X)$ are finite 
for an integer $p \geq 1$, then for any polynomial 
$H(x) = x^p + a_{p-1} x^{p-1} + \dots + a_0$,
$$
I^{(p)}(X) + \E\,H(X)^2 \geq 2 p!
$$
}

\vskip2mm
In particular, choosing $H(x) = x^p$, we get $I^{(p)}(X) \geq 2 p! - \E X^{2p}$.
Applying this to $\lambda X$ and optimizing over the parameter $\lambda>0$,
we arrive at the lower bound
$$
I^{(p)}(X) \, \E X^{2p} \geq \frac{1}{2}\, p!
$$

\vskip2mm
\section{{\bf Lower semi-continuity}}
\setcounter{equation}{0}

\vskip2mm
\noindent
We now consider the lower semi-continuity of the Fisher information.
In the case $p=1$, the next statement corresponds to Proposition 3.1 from
\cite{B-C-G}.

\vskip5mm
{\bf Proposition 7.1.} {\sl Let $(X_n)_{n \geq 1}$ be a sequence of
random variables, and let $X$ be a random variable such that $X_n \Rightarrow X$ 
weakly in distribution as $n \rightarrow \infty$. For any integer $p \geq 1$,
\be
I^{(p)}(X) \, \leq \, \liminf_{n \rightarrow \infty}\, I^{(p)}(X_n).
\en
}

\vskip2mm
Since the general case requires some modifications in the argument used for
$p=1$ (especially in the last steps), we include the proof below.

\vskip2mm
{\bf Proof.} Denote by $\mathfrak P_p$ the collection of all probability 
densities $f$ on the real line with finite Fisher information of order $p$, 
and let $\mathfrak P_p(I)$ denote the subset of all densities which have Fisher
information at most $I$. Since the case $p=1$ in (7.1) is known,
let $p \geq 2$.

For the proof of (7.1), we may assume that $I(X_n) \rightarrow I$ as 
$n \rightarrow \infty$ for some finite constant $I$. Then, for sufficiently 
large $n$, and without loss of generality for all $n \geq 1$, the random 
variables $X_n$ have densities $f_n$ belonging to $\mathfrak P_p(I+1)$. 
In particular, these densities have derivatives $f_n^{(k)}$ up to order 
$p-1$, such that the functions $f_n^{(p-1)}$ are absolutely continuous 
and have derivatives $f_n^{(p)}$ which are defined and finite almost everywhere. 

According to Proposition 4.1, for every $k = 0,1,\dots,p-1$,
\be
\|f_n^{(k)}\|_{\rm TV} + \sup_x |f_n^{(k)}(x)| < C_p (I+1)
\en
with a constant $C_p$ depending on $p$ only. By the second Helly theorem 
(cf. e.g. \cite{K-F}), $f_n^{(k)}(x)$ are convergent pointwise to some functions 
$g_k(x)$ of bounded total variation along a certain subsequence. 
For simplicity of notations, let this subsequence be a whole sequence, that is,
\be
\lim_{n \rightarrow \infty} f_n^{(k)}(x) = g_k(x) \quad 
{\rm for \ all} \ x \in \R.
\en
Due to (7.2), this property can be complemented by the $L^1$ convergence 
on bounded intervals (for a proof, cf. \cite{B3}, Proposition 11.4): For all $a<b$ ,
\be
\lim_{n \rightarrow \infty} \int_a^b |f_n^{(k)}(x) - g_k(x)|\,dx = 0.
\en

Putting $g_0 = g$, we have, in particular,
$\lim_{n \rightarrow \infty} f_n(x) = g(x)$ and
\be
\lim_{n \rightarrow \infty} \int_a^b |f_n(x) - g(x)|\,dx = 0, \quad
-\infty < a < b < \infty.
\en
Necessarily, $g(x) \geq 0$ and $\int_{-\infty}^\infty g(x)\,dx \leq 1$ 
(by Fatou's lemma). In fact, $\int_{-\infty}^\infty g(x)\,dx = 1$ which
follows from the weak convergence of the distributions of $X_n$.
Indeed, the latter implies and is actually equivalent to the property 
that, for any open set $G \subset \R$,
$$
\P(X \in G) \leq \liminf_{n \rightarrow \infty} \, \P(X_n \in G)
$$
(cf. e.g. \cite{Bil}). Given $\ep > 0$, choose an interval $G = (a,b)$
such that $\P(X \in G) > 1-\ep$, so that
$$
\liminf_{n \rightarrow \infty} \, \int_G f_n(x)\,dx > 1 -\ep.
$$
By (7.5), the last integrals are convergent to
$\int_G g(x)\,dx$. Therefore, $\int_G g(x)\,dx \geq 1-\ep$ and thus
$\int_{-\infty}^\infty g(x)\,dx \geq 1-\ep$ for any $\ep>0$.
Hence $g$ is a probability density. Since, the property (7.5) is stronger 
than the weak convergence, we also conclude that the distribution of $X$ 
is absolutely continuous with respect to the Lebesgue measure
and has density $g$.

If $1 \leq k \leq p-1$,  from (7.3)-(7.4) it follows that, for all $a,b \in \R$,
\be
\int_a^b g_k(x)\,dx = g_{k-1}(b) - g_{k-1}(a).
\en
This means that $g_k$ represents a Radon-Nikodym derivative of $g_{k-1}$.
In particular, $g_{k-1}$ is continuous, and we conclude that the density 
$g$ has $p-2$ continuous derivatives $g^{(k)} = g_k$, $1 \leq k \leq p-2$. 
The case $k = p-1$ in (7.6) similarly implies that $g_{p-1}$ 
is a Radon-Nikodym derivative of $g_{p-2} = g^{(p-2)}$.
Hence, $g^{(p-2)}$ is almost everywhere differentiable and has
a finite derivative $g_{p-1} = g^{(p-1)}$.

Now, by Proposition 3.3,
\be
\int_{-\infty}^\infty |f_n^{(p)}(x)|^2\,dx \leq C_p (I+1)^{3/2}.
\en
Since the unit ball of any separable $L^2$-space is weakly compact, 
there is a subsequence of $f_n^{(p)}$ which is weakly convergent 
to some function $g_p \in L^2(\R)$. For simplicity of notations, 
again let this subsequence be a whole sequence, so that
\be
\int_{-\infty}^\infty f_n^{(p)}(x) u(x)\,dx \rightarrow 
\int_{-\infty}^\infty g_p(x) u(x)\,dx \quad 
(n \rightarrow \infty)
\en
for any $u \in L^2(\R)$. Choosing here the indicator function $u = 1_{(a,b)}$ 
of a finite interval and applying (7.3) with $k = p-1$, we obtain that
$$
g^{(p-1)}(b) - g^{(p-1)}(a) = \int_a^b g_p(x)\,dx.
$$
This means that $g_p$ appears as a Radon-Nikodym derivative of $g_{p-1}$.
In particular, $g_{p-1}$ is continuous, and therefore $g$ has $p-1$ continuous
derivatives $g^{(k)} = g_k$, $1 \leq k \leq p-1$. Thus, the function $g$ belongs 
to the class $\mathfrak C^p$ with $g^{(p)} = g_p$ and
$I^{(p)}(X) = I^{(p)}(g)$.

Finally, consider the sequence of functions
$$
h_n(x,\lambda) = f_n^{(p)}(x)\,e^{-\lambda f_n(x)/2}, \quad x \in \R, \ 
\lambda>0.
$$
They have bounded $L^2$-norms on the half-plane $\R \times \R_+$, namely
\bee
\|h_n\|^2_{L^2(\R \times \R_+)} 
 & = &
\int_{-\infty}^\infty \int_0^\infty h_n(x,\lambda)^2\, dx\,d\lambda \\
 & = &
\int_{f_n(x)>0} \frac{f_n^{(p)}(x)^2}{f_n(x)}\,dx \, = \, I^{(p)}(X_n) 
 \, \leq \, I+1.
\ene
Here we applied Proposition 3.1, according to which $f_n^{(p)}(x) = 0$
for almost all $x$ on the set where $f_n(x) = 0$. 

Let us verify that $h_n$ are weakly convergent in $L^2$ to the function
$$
h(x,\lambda) = g^{(p)}(x)\,e^{-\lambda g(x)/2}
$$
on every rectangle $R = [-T,T] \times [\lambda_0,\lambda_1]$
with fixed $T>0$ and $\lambda_1 > \lambda_0 > 0$. Write
\begin{eqnarray}
h_n(x,\lambda) - h(x,\lambda) 
 & = &
f_n^{(p)}(x)\,\big(e^{-\lambda f_n(x)/2} - e^{-\lambda g(x)/2}\big) \nonumber \\
 & & + \
\big(f_n^{(p)}(x) - g^{(p)}(x)\big)\,e^{-\lambda g(x)/2}.
\end{eqnarray}
Given a Borel measurable function $u \in L^2(\R \times \R_+)$ supported 
on $R$, define
$$
u_1(x) = \int_{\lambda_0}^{\lambda_1} e^{-\lambda g(x)/2}\,u(x,\lambda)\,d\lambda,
\quad x \in \R.
$$
It is Borel measurable, supported on $[-T,T]$, and is bounded, since 
$g$ is continuous (hence bounded
on $[-T,T]$). Therefore, by the Fubini theorem and the weak convergence (7.8),
\begin{eqnarray}
\int\!\!\!\int_R \big(f_n^{(p)}(x) - g^{(p)}(x)\big)\,e^{-\lambda g(x)/2}\,
u(x,\lambda)\,dx\, d\lambda & & \nonumber \\
 & & \hskip-55mm = \
\int_{-\infty}^\infty \big(f_n^{(p)}(x) - g_p(x)\big)\, u_1(x)\,dx \, \rightarrow 0 \,
\qquad (n \rightarrow \infty).
\end{eqnarray}

Next, by (7.3) with $k=0$, we have 
$f_n(0) \rightarrow g(0)$ as $n \rightarrow \infty$. Using the representation
$$
(f_n(x) - g(x)) - (f_n(0) - g(0)) = \int_0^x (f_n'(y) - g'(y))\,dy,
$$
from (7.4) with $k=1$ it also follows that
$$
\ep_n = \sup_{|x| \leq T} |f_n(x) - g(x)| \rightarrow 0.
$$
Hence
$$
|e^{-\lambda f_n(x)/2} - e^{-\lambda g(x)/2}| \leq C\ep_n
$$
with some constant $C$ (which may depend on $T$ and $\lambda_j$). 
Using Cauchy's inequality, this gives
\bee
\Big|\int\!\!\!\int_R f_n^{(p)}(x)\,
\big(e^{-\lambda f_n(x)/2} - e^{-\lambda g(x)/2}\big)\,
u(x,\lambda)\,dx\, d\lambda\Big|^2 & & \\
 & & \hskip-70mm \leq \ (C\ep_n)^2\,(\lambda_1 - \lambda_0)
\int_{-\infty}^\infty f_n^{(p)}(x)^2\,dx  
\int\!\!\!\int_R u(x,\lambda)^2\,dx\, d\lambda \, \rightarrow \, 0
\ene
as $n \rightarrow \infty$, thanks to (7.7) in the last step.
Combining this with (7.10) and returning to (7.9), we conclude that
$$
\int\!\!\!\int_R (h_n(x,\lambda) - h(x,\lambda))\,
u(x,\lambda)\,dx\, d\lambda \, \rightarrow 0 \, \quad (n \rightarrow \infty),
$$
which means that $h_n$ is weakly convergent to $h$ in the space
$L^2(R)$. Therefore
$$
\|h\|_{L^2(R)}^2 \, \leq \, \liminf_{n \rightarrow \infty}\, \|h_n\|_{L^2(R)}^2 
 \, \leq \, \liminf_{n \rightarrow \infty} I^{(p)}(X_n) \, = \, I.
$$
Thus,
$$
\int_{-T}^T \int_{\lambda_0}^{\lambda_1} 
g^{(p)}(x)^2\,e^{-\lambda g(x)}\,dx\, d\lambda \leq I.
$$
Letting here $T \rightarrow \infty$, $\lambda_1 \rightarrow \infty$ and
$\lambda_0 \rightarrow 0$, we arrive at (7.1).
\qed

\vskip5mm
{\bf Remark 7.2.} On the set $\mathfrak P_p(I)$ the weak convergence of the
associated probability distributions coincides with the convergence in total 
variation distance (which corresponds to the convergence of probability
densities in the $L^1$-norm). For the proof, suppose that
$X_n \Rightarrow X$ weakly in distribution as $n \rightarrow \infty$ with
$I^{(p)}(X_n) \leq I$. Then $X_n$ have densities $f_n$ of class $\mathfrak C^p$
with $I^{(p)}(f_n) \leq I$. We need to show that $X$ has a density $f$ 
in the same class such that
\be
\int_{-\infty}^\infty |f_n(x) - f(x)|\,dx \rightarrow 0 \quad
(n \rightarrow \infty).
\en
Equivalently, it is sufficient to show that from any prescribed subsequence 
$f_{n_k}$ one may extract a further subsequence $f_{n_{k_l}}$ which is
convergent in $L^1$ to some density $f$. Arguing as in the beginning
of the proof of Proposition 7.1, we obtain such a subsequence
with the property that $f_{n_{k_l}}(x) \rightarrow f(x)$ for all
$x \in \R$ as $l \rightarrow \infty$ for some density $f$.
Applying Scheffe's lemma, this leads to (7.11) for $f_{n_{k_l}}$ and $f$.

\vskip7mm
\section{{\bf Convex mixtures of probability measures}}
\setcounter{equation}{0}

\vskip2mm
\noindent
We will consider some properties of the Fisher-type information
for random variables whose distributions are representable in a natural way
as mixtures of probability measures (including convolutions). 
In order to make all statements rigorous and as general as possible, 
first let us give a few remarks about the notion of mixture.

Denote by $\mathfrak M$ the collection of all probability measures on the real
line. We treat it as a separable metric space with the topology of weak convergence 
which may be metrized using the L\'evy distance, for example.
As such, this space has a canonical Borel $\sigma$-algebra generated by 
the collection of all open subsets of $\mathfrak M$.

\vskip5mm
{\bf Lemma 8.1.} {\sl For any Borel set $A \subset \R$, the functional 
$T_A(\nu) = \nu(A)$ is Borel measurable on $\mathfrak M$. Moreover,
the functional
$$
T_u(\nu) = \int_{-\infty}^\infty u\,d\nu
$$
is Borel measurble on $\mathfrak M$, whenever the function
$u:\R \rightarrow \R$ is bounded and Borel measurable. 
}

\vskip5mm
{\bf Proof.} Consider the collection $\mathfrak A$ of all Borel sets $A \subset \R$ 
such that $T_A$ is Borel measurable on $\mathfrak M$. Let us list several basic properties
of this functional.

\vskip2mm
1) \ For the union $A$ of disjoint Borel sets $A_n$, we have
$T_A = \sum_{n=1}^\infty T_{A_n}$.

2) \ For the monotone limit $A$ of increasing or decreasing Borel sets $A_n$, 
$T_A = \lim_{n \rightarrow \infty} T_{A_n}$.

3) \ For the complement $\bar A = \R \setminus A$, we have
$T_{\bar A} = 1 - T_A$.

4) \ More generally, $T_{A \setminus B} = T_A - T_B$ as long as $B \subset A$. 

5) \ If $A$ is closed, and 
$\nu_n \rightarrow \nu$ weakly in $\mathfrak M$, then
$$
\limsup_{n \rightarrow \infty} \, \nu_n(A) \leq \nu(A).
$$
The last property is equivalent to saying that the functional $T_A$ is upper 
semi-continuous on $\mathfrak M$. Hence, it is Borel measurable on $\mathfrak M$, 
that is, $A \in \mathfrak A$. Thus, $\mathfrak A$ is a monotone class containing
all semi-open intervals $(a,b] = (-\infty,b] \setminus (-\infty,a]$, and
therefore, this class contains all Borel subsets of the real line.

For the second assertion, first note that if $u$ is simple in the sense 
that it is a finite linear combination of indicator functions $1_A$ of Borel sets
$A \subset \R$, we are reduced to the previous step. 
In the general case, if $|u| \leq M$, there is a sequence
of simple functions $u_n$ with values in $[-M,M]$ such that
$u_n(x) \rightarrow u(x)$ for all $x \in \R$ as $n \rightarrow \infty$.
By the Lebesgue dominated convergence theorem,
$T_{u_n}(\nu) \rightarrow T_u(\nu)$ for any $\nu \in \mathfrak M$, implying that
$T_u$ is Borel measurable on $\mathfrak M$.
\qed

\vskip2mm
Lemma 8.1 justifies the following:

\vskip4mm
{\bf Definition 8.2.} Let $\pi$ be a Borel probability measure on the space
$\mathfrak M$. A convex mixture of probability measures on the real line 
with a mixing measure $\pi$ is given by
\be
\mu(A) = \int_{\mathfrak M} \nu(A)\,d\pi(\nu), \quad A \subset \R \ ({\rm Borel}).
\en

\vskip2mm
Recall that in the space $\mathfrak M$ there is a canonical metric defined 
by the total variation distance $\|\nu - \lambda\|_{\rm TV}$ between 
probability measures. It generates a stronger topology, and $\mathfrak M$ 
is not separable with respect to this metric (because, for example, 
$\|\delta_x - \delta_y\|_{\rm TV} = 2$ for all $x,y \in \R$, $x \neq y$).
Nevertheless, the balls for this metric are Borel measurable 
for the weak topology. Indeed, for any signed Borel measure 
$\nu_0$ on $\R$,
$$
\|\nu - \nu_0\|_{\rm TV} = \sup_u \, |T_u(\nu) - T_u(\nu_0)|,
$$
where the supremum may be taken over the set $C_0$ of all continuous, 
compactly supported functions $u$ on $\R$ such that $|u| \leq 1$. Moreover, 
this supremum can be restricted to a countable set, since the space 
$C_0$ is separable for the supremum-norm. Since for each $u$ in $C_0$, 
the functional $\nu \rightarrow T_u(\nu)$ is continuous on $\mathfrak M$, 
the functional $\nu \rightarrow \|\nu - \nu_0\|_{\rm TV}$ 
is Borel measurable.

\vskip5mm
{\bf Lemma 8.3.} {\sl The collection $\mathfrak M_0$ of all absolutely 
continuous probability measures on the real line (with respect to 
the Lebesgue measure) represents a Borel set in $\mathfrak M$.
}

\vskip5mm
{\bf Proof.} Recall the following well-known characterization:
A Borel probability measure $\nu$ on the real line with distributions function 
$F(x) = \nu((-\infty,x])$, $x \in \R$, is absolutely continuous (with respect to 
the Lebesgue measure), if and only if, for any $\ep>0$, there is $\delta > 0$,
such that, for any finite collection of non-overlapping intervals
$(a_i,b_i) \subset \R$, $1 \leq i \leq n$,
$$
\sum_{i=1}^n (b_i - a_i) < \delta \ \Longrightarrow \
\sum_{i=1}^n (F(b_i) - F(a_i)) < \ep.
$$
Since $F$ is non-decreasing and right-continuous, here one may additionally 
require that the endpoints $a_i$ and $b_i$ represent rational numbers. 
Also, one may replace open intervals in this definition with semi-open intervals 
$(a_i,b_i]$, leading to the increments $F(b_i) - F(a_i-)$. Define 
$$
\mathfrak A = \Big\{A = \bigcup_{i=1}^n\, (a_i,b_i]: 
a_1 < b_1 \leq \dots \leq a_n < b_n, \ a_i, b_i \in \mathbb Q, \ n \geq 1\Big\}
$$
and rewrite the definition of the absolute continuity of $\nu$ as the property 
that, for any $\ep>0$, there is $\delta > 0$ such that, for any $A \in \mathfrak A$, 
$\lambda(A) \leq \delta \Rightarrow \nu(A) \leq \ep$, where $\lambda$ 
denotes the Lebesgue measure on the real line. In terms of the functional
$$
M_\delta(\nu) = \sup\big\{\nu(A): 
A \in \mathfrak A, \ \lambda(A) \leq \delta\big\},
$$
this is equivalent to saying that
$$
M(\nu) \, \equiv \, \inf_{\delta > 0}\, M_\delta(\nu) \, = \, 
\inf_k\, M_{1/k}(\nu) \, = \, 0.
$$

Now, the crucial point is that the collection $\mathfrak A$ is countable.
Applying Lemma 8.1, we conclude that every functional $M_\delta$ is Borel
measurable on $\mathfrak M$ as the supremum of countably many Borel 
measurable functionals. Therefore, $M$ is Borel measurable as well, and 
the set $\mathfrak M_0$ is described as the pre-image $M^{-1}(\{0\})$.
\qed

\vskip5mm
Denote by $\mathfrak P$ the collection of all (probability) densities $f$
on the real line. It is a closed convex subset of $L^1(\R)$ with respect 
to the usual $L^1$-metric. With every $f$ in $\mathfrak P$ we associate 
the probability measure $\mu_f$ with this density. By Lemma 8.3, the collection 
$\mathfrak M_0 = \{\mu_f: f \in \mathfrak P\}$ represents a Borel set in 
$\mathfrak M$. One can thus identify $\mathfrak P$ and $\mathfrak M_0$ 
by means of the bijective map $f \rightarrow \mu_f$.

\vskip3mm
{\bf Lemma 8.4.} {\sl The Borel $\sigma$-algebra in $\mathfrak P$ induced 
from $L^1(\R)$ coincides with the Borel $\sigma$-algebra in $\mathfrak M_0$ 
induced from $\mathfrak M$.
}

\vskip3mm
{\bf Proof.} Given a sequence $f_n$ and $f$ in $\mathfrak P$, the weak 
convergence $\mu_{f_n} \rightarrow \mu_f$ in $\mathfrak M$ is equivalent to
$$
\int_{-\infty}^x f_n(y)\,dy \rightarrow \int_{-\infty}^x f(y)\,dy \quad
{\rm for \ any} \ x \in \R
$$
(and actually uniformly over all $x$). It is weaker than the convergence in $L^1$
$$
\|\mu_{f_n} - \mu_f\|_{\rm TV} = \|f_n - f\|_1 =
\int_{-\infty}^\infty |f_n(y)-f(y)|\,dy \, \rightarrow \, 0,
$$
which is equivalent to the convergence of the measures in total variation distance.
Hence, the Borel $\sigma$-algebra in $\mathfrak M_0$ induced from $\mathfrak M$
is (formally) smaller than the Borel $\sigma$-algebra in $\mathfrak P$ induced 
from $L^1(\R)$, using the identification of $\mathfrak P$ and $\mathfrak M_0$.
 
For the opposite inclusion, first recall that the Borel $\sigma$-algebra 
in $L^1(\R)$ is generated by the $L^1$-balls 
$$
B = \{f \in L^1(\R): \|f - f_0\|_1 < r\}, \quad f_0 \in L^1, \ r>0
$$
(since the space $L^1$ is separable). Hence, it is sufficient to see that 
any set of the form $D = B \cap \mathfrak P$ is Borel measurable in 
$\mathfrak M$ (where we use Lemma 8.3). This is the same as saying that
the balls in $\mathfrak M$ for the total variation distance are 
Borel measurable, which has been already explained.
\qed

\vskip2mm
As a consequence from Lemma 8.4, one may use Definition 8.2 starting from
a Borel probability measure $\pi$ on $\mathfrak P$. Following this definition,
one can define the convex mixture according to (8.1):
\be
\mu(A) = \int_{\mathfrak P} \bigg[\int_A g(x)\,dx\bigg]\,d\pi(g), \quad 
A \subset \R \ ({\rm Borel}).
\en
This measure belongs to $\mathfrak M_0$ and has some density 
$f(x) = \frac{d\mu(x)}{dx}$ called the (convex) mixture of densities 
with mixing measure $\pi$. For short,
$$
f(x) = \int_{\mathfrak P} g(x)\,d\pi(g), \quad x \in \R.
$$

\vskip5mm
\section{{\bf Convexity of the Fisher-type information}}
\setcounter{equation}{0}

\vskip2mm
\noindent
Another general property of the Fisher-type information is its convexity.

\vskip5mm
{\bf Proposition 9.1.} {\sl Given probability densities $f_i$ on the real line 
and weights $\alpha_i > 0$ such that $\sum_{i=1}^n \alpha_i = 1$, we have
\be
I^{(p)}(f) \leq \sum_{i=1}^n \alpha_i I^{(p)}(f_i), \quad {\sl where} \ 
f = \sum_{i=1}^n \alpha_i f_i.
\en
}

{\bf Proof.} This follows from the fact that the function $R(u,v) = u^2/v$ 
is 1-homogeneous and convex on the upper half-plane $u \in \R$, $v > 0$. 
For more details,
one may assume that $n=2$ and $I^{(p)}(f_i) < \infty$, $i = 1,2$. Thus, 
$f_1$, $f_2$ and $f$ belong to the class $\mathfrak C^p$ with 
$f^{(p)} = \alpha_1 f_1^{(p)} + \alpha_2 f_2^{(p)}$. Let $G$ denote
the set of all points $x \in \R$ where $f(x) > 0$ and such that the derivatives 
$f_i^{(p-1)}(x)$ are differentiable at $x$, so that
$$
I^{(p)}(f) = \int_G R\big(f^{(p)}(x),f(x)\big)\,dx.
$$

The set $G$ can be decomposed into the three measurable parts
\bee
G_0 
 & = &
\{x \in G: f_1(x) > 0, \ f_2(x) > 0\}, \\
G_1 
 & = &
\{x \in G: f_1(x) > 0, \ f_2(x) = 0\}, \\
G_2 
 & = &
\{x \in G: f_1(x) = 0, \ f_2(x) > 0\}.
\ene
On the first part, due to the convexity of $R$,
$$
\int_{G_0} R\big(f^{(p)}(x),f(x)\big)\,dx \, \leq \,
\alpha_1 \int_{G_0} R\big(f_1^{(p)}(x),f_1(x)\big)\,dx +
\alpha_2 \int_{G_0} R\big(f_2^{(p)}(x),f_2(x)\big)\,dx.
$$
If $x \in G_1$, then $f(x) = \alpha_1 f_1(x)$ and
$$
\int_{G_1} R\big(f^{(p)}(x),f(x)\big)\,dx \, = \,
\alpha_1 \int_{G_1} R\big(f_1^{(p)}(x),f_1(x)\big)\,dx.
$$
Similarly,
$$
\int_{G_2} R\big(f^{(p)}(x),f(x)\big)\,dx \, = \,
\alpha_2 \int_{G_2} R\big(f_2^{(p)}(x),f_2(x)\big)\,dx.
$$
Summing the last inequality with the last two equalities,
we obtain (9.1).
\qed

\vskip5mm
As a consequence of Propositions 9.1 and 7.1, we obtain:

\vskip5mm
{\bf Corollary 9.2.} {\sl Given a number $I>0$, the collection $\mathfrak P_p(I)$ 
of all probability densities $f$ on the real line with $I^{(p)}(f) \leq I$ represents
a convex closed set in $L^1(\R)$.
}

\vskip5mm
Here, the closeness is understood with respect to the $L^1$-distance, but
may be also understood with respect to the weak convergence of
measures associated to probability densities (as explained in Remark 7.2).

We need to extend Jensen’s inequality (9.1) to arbitrary ``continuous” 
convex mixtures of densities and probability distributions. For this aim, 
we temporarily employ the notation $I^{(p)}(\mu)$ for $I^{(p)}(X)$, 
when a random variable $X$ is distributed according to $\mu$.

\vskip5mm
{\bf Proposition 9.3.} {\sl If a probability density $f$ is a convex mixture 
of densities with mixing measure $\pi$ on $\mathfrak P$, then
\be
I^{(p)}(f) \leq \int_{\mathfrak P} I^{(p)}(g)\,d\pi(g).
\en
More generally, if a probability measure $\mu$ is a convex mixture 
of probability measures with mixing measure $\pi$ on $\mathfrak M$, then
\be
I^{(p)}(\mu) \leq \int_{\mathfrak M} I^{(p)}(\nu)\,d\pi(\nu).
\en
}

The integrals in (9.2)-(9.3) make sense, since the functionals 
$g \rightarrow I^{(p)}(g)$ and $\nu \rightarrow I^{(p)}(\nu)$ are
lower semi-continuous and hence Borel measurable on $\mathfrak P$ and
$\mathfrak M$, respectively (Proposition 7.1 and Lemma 8.3). 

\vskip2mm
{\bf Proof.} 
To derive (9.3), one may assume that the integral on the right-hand
side is finite. But then $I^{(p)}(\nu)$ is finite for $\pi$-almost all $\nu$,
which implies that the measure $\pi$ is supported on $\mathfrak M_0$. 
In this case, $\mu$ belongs to $\mathfrak M_0$, and (9.3) is reduced to (9.2).

The proof of the inequality (9.2) is similar to the one of Proposition 3.3
in \cite{B-C-G} for the case $p=1$, and here we provide details with slight modifications.
We may assume that the integral in (9.2) is finite, so that $\pi$ is supported on the convex 
(Borel measurable) set $\mathfrak P_p = \cup_I \mathfrak P_p(I)$. 

{\it Step} 1. Suppose that the measure $\pi$ is supported on some convex 
compact set  $K$ contained in $\mathfrak P_p(I)$. Since the functional 
$g \rightarrow I^{(p)}(g)$ is finite, convex and lower semi-continuous on $K$, 
it admits the representation
$$
I^{(p)}(g) \, = \, \sup_{l \in L} \, l(g), \quad g \in K,
$$
where $L$ is the family of all continuous affine 
functionals $l$ on $L^1(\R)$ such that $l(g) < I^{(p)}(f)$ for all $g \in K$ 
(cf. \cite{Me}, Theorem T7, for a more general setting of
locally convex spaces). Being restricted to probability densities, any such functional
has the form $l(g) =  \int_{-\infty}^\infty g(x)\psi(x)\,dx$ for some measurable
function $\psi$. Hence
$$
I^{(p)}(g) \, = \, \sup_{\psi \in \Psi} \, \int_{-\infty}^\infty g(x)\psi(x)\,dx
$$
for some family $\Psi$ of bounded measurable functions $\psi$ on $\R$. 
As a consequence, using the definition (8.2) for the measure $\mu$ with density $f$
and applying Fubini's theorem, we get
\bee
\int_{\mathfrak P} I^{(p)}(g)\,d\pi(g)
 & \geq &
\sup_{\psi \in \Psi} \ \int_{\mathfrak P} \bigg[
\int_{-\infty}^\infty g(x)\psi(x)\,dx\bigg]\,d\pi(g) \\
 & = &
\sup_{\psi \in \Psi} \ \int_{-\infty}^\infty g(x)\psi(x)\,dx
 \ = \ I^{(p)}(f),
\ene
which is the desired inequality (9.2).

{\it Step} 2. Suppose that $\pi$ is supported on $\mathfrak P_p(I)$
for some $I>0$. Since any finite Borel measure on $L^1(\R)$ is Radon, and 
since the set $\mathfrak P_p(I)$ is closed and convex, there is an increasing 
sequence of compact subsets $K_n \subset \mathfrak P_p(I)$ such that 
$\pi(\cup_n K_n) = 1$. Moreover, $K_n$ can be chosen to be convex 
(since the closure of the convex hull will be compact as well). 
Let $\pi_n$ denote the normalized restriction of $\pi$ to $K_n$ 
with sufficiently large $n$ so that $c_n = \pi(K_n) > 0$, 
and define its baricenter
\be
f_n(x) = \int_{K_n} g(x)\,d\pi_n(g).
\en
Since $\|\pi_n - \pi\|_{\rm TV} \leq 2(1-c_n)$,
from (8.2) it follows that the measures $\mu_n$ with densities $f_n$ satisfy
$|\mu_n(A) - \mu(A)| \leq 2(1-c_n)$ for any Borel set $A \subset \R$. Hence
$$
\|\mu_n - \mu\|_{\rm TV} = \|f_n - f\|_1 \leq 4(1-c_n) \rightarrow 0,
$$ 
and the relation (7.1) holds: 
$I^{(p)}(f) \leq \liminf_{n \rightarrow \infty} \, I^{(p)}(f_n)$.
On the other hand, by the previous step and the monotone convergence
theorem,
\begin{eqnarray}
I^{(p)}(f_n) 
 & \leq & 
\int_{K_n} I^{(p)}(g)\,d\pi_n(g) \nonumber\\
 & = &
\frac{1}{c_n}\, \int_{K_n} I^{(p)}(g)\,d\pi(g) \, \rightarrow 
\int_{\mathfrak P_p(I)} I^{(p)}(g)\,d\pi(g),
\end{eqnarray}
and we obtain (9.2).

{\it Step} 3. In the general case, we may apply Step 2 to 
the normalized restrictions $\pi_n$ of $\pi$ to the sets 
$K_n = \mathfrak P_p(n)$. Again, for the densities $f_n$ defined 
in (9.4), we obtain (9.5), where $\mathfrak P_p(I)$ should be 
replaced with $\mathfrak P_p$. Another application of the lower
semi-continuity of the Fisher information finishes the proof.
\qed

\vskip7mm
\section{{\bf Monotonicity and continuity along convolutions}}
\setcounter{equation}{0}

\vskip2mm
\noindent
As a consequence of Proposition 9.3, the functional $I^{(p)}$ is monotone
under convolutions.

\vskip5mm
{\bf Proposition 10.1.} {\sl For all independent random variables $X$ and $Z$,
\be
I^{(p)}(X + Z) \leq I^{(p)}(X).
\en
}

{\bf Proof.}
Let $\nu$ denote the distribution of $X$, and let $\nu_z(A) = \nu(A - z)$ 
be the shift of $\nu$ ($z \in \R$). The distribution
of $X+Z$ represents the mixture
$$
\mu = \int_{-\infty}^\infty \nu_z\,dP(z),
$$
where $P$ is the distribution of $Z$. The map $T:\R \rightarrow \mathfrak M$
defined by $T(z) = \nu_z$ is continuous, so, the image $B = T(\R)$ is 
a $\sigma$-compact, hence a Borel set in $\mathfrak M$. This map pushes 
forward $P$ to a Borel probability measure $\pi$ supported on $B$. It remains 
to apply (9.3) and note that $I^{(p)}(\nu_z) = I^{(p)}(\nu)$ for all $z$.
\qed

\vskip5mm
Combining Propositions 7.1 and 10.1, we obtain the continuity
property of the functional $I^{(p)}$ for convolved densities as stated in
Theorem 1.4: For all independent random variables $X$ and $Z$,
\be
\lim_{\ep \rightarrow 0} \, I^{(p)}(X + \ep Z) = I^{(p)}(X).
\en

\vskip2mm
{\bf Proof of Theorem 1.4.}
The distributions of $X + \ep Z$ are weakly convergent to the distribution
of $X$ as $\ep \rightarrow 0$, so that, by (7.1),
$$
I^{(p)}(X) \leq \liminf_{\ep \rightarrow 0} \, I^{(p)}(X + \ep Z).
$$
On the other hand, $I^{(p)}(X + \ep Z) \leq I^{(p)}(X)$, by (10.1).
Both inequalities lead to (10.2).
\qed

\vskip5mm
{\bf Corollary 10.2.} {\sl Suppose that a normal random variable $Z$
is independent of the random variable $X$. Then the function
$\ep \rightarrow I^{(p)}(X + \ep Z)$ is finite and non-decreasing
in $\ep > 0$.
}

\vskip5mm
Indeed, let $Z \sim N(0,1)$. By Proposition 10.1 and according to Example 2.1,
$$
I^{(p)}(X + \ep Z) \leq I^{(p)}(\ep Z) = p!\,\ep^{-2p}.
$$
The monotonicity follows from the fact that the convolution of
Gaussian measures is Gaussian.

\vskip5mm
{\bf Remark 10.3.} The functional
$$
I_p(X) = I_p(f) = \E\,|\rho(X)|^p = 
\int_{-\infty}^\infty \Big|\frac{f'(x)}{f(x)}\Big|^p\,f(x)\,dx
$$
satisfies similar properties as the Fisher information
(in the case $p=1$), such as the lower semi-continuity and the monotonicity
$$
I_p(X+Y) \leq \min\{I_p(X),I_p(Y)\},
$$
which holds true for all for independent summands $X$ and $Y$. Hence, 
similarly to Corollary 10.2, $I_p(X + \ep Z) < \infty$ for all $p$, assuming 
that $X$ and $Z$ are independent, and $Z \sim N(0,1)$.
As another consequence, we have the analog of (10.2)
\be
\lim_{\ep \rightarrow 0} \, I_p(X + \ep Z) = I_p(X).
\en

It is shown in \cite{B2} that, if $p \geq 1$ is an integer and the random 
variables $(X_i)_{1 \leq i \leq p+1}$ are independent
and have densities with finite total variation $b_i = I_1(X_i)$, then
$$
I_p(X_1 + \dots + X_{p+1}) \leq c_p\, b_1 \dots b_{p+1}\,
\Big(\frac{1}{b_1} + \dots + \frac{1}{b_{p+1}}\Big)
$$
with constant $c_p = p^p/(2^p p!)$.

\vskip7mm
\section{{\bf Representations in terms of isoperimetric profile}}
\setcounter{equation}{0}

\vskip2mm
\noindent
If a continuous probability density $f$ is supported and positive on 
the interval $(a,b) \subset \R$, finite or not, the associated distribution
may be characterized, up to a shift parameter, by the function 
\be
L(t) = f(F^{-1}(t)), \quad 0 < t < 1,
\en
called sometimes the isoperimetric profile. This follows from the equality
$$
F^{-1}(t_2) - F^{-1}(t_1) = \int_{t_1}^{t_2} \frac{dt}{L(t)}, \quad 
0 < t_1,t_2 < 1.
$$
Here 
$F^{-1}:(0,1) \rightarrow (a,b)$ denotes the inverse of the distribution 
function $F(x) = \int_a^x f(y)\,dy$ restricted to $(a,b)$. The definition (11.1)
may be written equivalently as
\be
f(x) = L(F(x)), \quad a < x < b.
\en

If $f$ is locally absolutely continuous on $(a,b)$ and has derivative $f'$, 
both $F$ and $F^{-1}$ will be $C^1$-smooth functions with absolutely 
continuous derivatives. Hence, $L$ is also
locally absolutely continuous on $(0,1)$. Differentiating (11.2), we obtain 
$f' = L'(F) f$ a.e. in $(a,b)$, implying that the random variable $X$ 
with density $f$ has the Fisher information
$$
I(X) = \int_a^b L'(F(x))^2 f(x)\,dx = \int_0^1 L'(t)^2\,dt.
$$
More generally, the moments of the scores of $X$ are given by
\be
I_p(X) = \int_0^1 |L'(t)|^p\,dt.
\en

If $f'$ is locally absolutely continuous on $(a,b)$ and has derivative $f''$, 
then both $F$ and $F^{-1}$ are $C^2$-smooth with absolutely continuous 
second order derivatives. Hence, $L$ has a locally absolutely continuous 
derivative $L'$ whose derivative $L''$ is defined and finite a.e.
on $(0,1)$. Starting from (11.1), we get
$(L^2)' =  2f'(F^{-1})$ and $(L^2)'' =  2f''(F^{-1})/f(F^{-1})$.
This gives:

\vskip5mm
{\bf Proposition 11.1.} {\sl Suppose that the density $f$ of the random 
variable $X$ is supported and positive on an interval,  finite or not. 
If it is of the class $\mathfrak C^1$ or $\mathfrak C^2$, then respectively
\begin{eqnarray}
I(X)
 & = &
\int_0^1 L'(t)^2\,dt, \nonumber \\
I^{(2)}(X) 
 & = &
\frac{1}{4} \int_0^1 \big(L^2(t)''\big)^2\,dt \ = \
\int_0^1 \big(L'(t)^2 + L(t) L''(t)\big)^2\,dt.
\end{eqnarray}
}

Note that, if $I^{(2)}(X)$ is finite, then necessarily
$$
\int_0^1 \big(L'(t)^2 + L(t) L''(t)\big)\,dt = 
\int_0^1 \big(L(t) L'(t)\big)'\,dt = 0.
$$
This follows from the property $f'(a+) = f'(b-) = 0$ which is
emphasized in Proposition 4.1. Indeed, using $L L' = f'(F^{-1})$, we get
$$
\int_{t_0}^{t_1} \big(L(t) L'(t)\big)'\,dt = 
L(t_1) L'(t_1) - L(t_0) L'(t_0) \rightarrow 0 \quad
{\rm as} \ \ t_0 \downarrow 0, \ t_1 \uparrow 1.
$$

There is another representation for the integral in (11.4).

\vskip5mm
{\bf Proposition 11.2.} {\sl Suppose that the density $f \in \mathfrak C^2$ 
of the random variable $X$ is supported and positive on an interval, 
finite or not. Then
\be
I^{(2)}(X) = \int_0^1 \Big(L''(t)^2 L(t)^2 + \frac{1}{3}\, L'(t)^4\Big)\, dt,
\en
as long as the latter integral is finite, which is equivalent to the finiteness of 
$I^{(2)}(X)$.
}

\vskip5mm
{\bf Proof.} By (11.4),
\be
I^{(2)}(X) = \int_0^1 \Big(L''(t)^2 L(t)^2 + L'(t)^4 + 
2 L''(t) L'(t)^2 L(t)\Big)\, dt.
\en
Integrating by parts, we have, for all $0 < t_0 < t_1 < 1$,
\bee
\int_{t_0}^{t_1} L''(t) L'(t)^2 L(t)\,dt 
 & = &
\int_{t_0}^{t_1} L'(t)^2 L(t)\,dL'(t) \\
 & & \hskip-25mm = \
L(t_1) L'(t_1)^3 - L(t_0) L'(t_0)^3 - 
\int_{t_0}^{t_1} L'(t)\,d\,(L'(t)^2 L(t)) \\
 & & \hskip-25mm = \
L(t_1) L'(t_1)^3 - L(t_0) L'(t_0)^3 - 
\int_{t_0}^{t_1} L'(t)^4\,dt - 2\int_{t_0}^{t_1} L''(t) L'(t)^2 L(t)\,dt.
\ene
Equivalently,
$$
3 \int_{t_0}^{t_1} L''(t) L'(t)^2 L(t)\,dt \, = \,
L(t_1) L'(t_1)^3 - L(t_0) L'(t_0)^3 - \int_{t_0}^{t_1} L'(t)^4\,dt.
$$
If we show that
\be
L(t) L'(t)^3 \rightarrow 0 \quad {\rm as} \ 
t \rightarrow 0 \ {\rm or} \ t \rightarrow 1, 
\en
then in the limit as $t_0 \rightarrow 0$ and $t_1 \rightarrow 1$
we would obtain that
$$
\int_0^1 L''(t) L'(t)^2 L(t)\, dt = -\frac{1}{3} \int_0^1 L'(t)^4\, dt.
$$
As a result, (11.6) will be simplified to the required representation (11.5).
Note that (11.7) is equivalent to the property
$\frac{f'(x)^3}{f(x)^2} \rightarrow 0$ as $x \rightarrow a$ and 
$x \rightarrow b$. 

In order to verify (11.7), we apply the Cauchy inequality and
use the assumption about the finiteness of the integral in (11.5) to get
$$
\Big(\int_0^1 L'(t)^2\,L(t)\, |L''(t)|\, dt\Big)^2 \leq
\int_0^1 L'(t)^4\, dt \int_0^1 L(t)^2 L''(t)^2\,dt < \infty.
$$
This implies that the function $u = L L'^3$
has a bounded total variation on $(0,1)$. Indeed, its derivative
$$
u' = L'^4 + 3 L L'^2 L''
$$
is integrable. Therefore, the limits
$c_0 = u(0+)$ and $c_1 = u(1-)$ exist and are finite. Let us show that
necessarily $c_0 = c_1 = 0$. Suppose that $c_0 \neq 0$. We have
$$
(L(t)^{4/3})' = \frac{4}{3}\, L(t)^{1/3} L'(t) = 
\frac{4}{3}\, u(t)^{1/3} \, \rightarrow \, \frac{4}{3}\,c_0^{1/3}
$$
as $t \rightarrow 0$. Since $L(0+) = 0$ and $L(t)>0$ for $t \in (0,1)$, 
this implies that $c_0>0$ and moreover
$L^{4/3}(t) \leq 2 c_0^{1/3} t$ for all $t$ small enough, $0 < t \leq t_0$,
that is, $L(t) \leq (8c_0)^{1/4}\,t^{3/4}$. This gives
$$
L'(t) \sim \frac{c_0^{1/3}}{L(t)^{1/3}} \geq \frac{c'}{t^{1/4}}, \quad 0 < t \leq t_0,
$$
with some constant $c'>0$. As a consequence, the function $L'^4$ 
is not integrable on this interval, which contradicts to the assumption. 
Hence, necessarily $c_0 = 0$, and by a similar argument, $c_1 = 0$
as well. Thus, (11.7) is fulfilled.
\qed

\vskip7mm
\section{{\bf Lower bounds for $I^{(2)}$ in terms of $I_4$ and $I$}}
\setcounter{equation}{0}

\vskip2mm
\noindent
The representation (11.5) may be used for the lower bound on $I^{(2)}$ 
in terms of $I_4$ and $I$, in order to derive the relation (1.5) of Theorem 1.3:
\be
I^{(2)}(X) \geq \frac{1}{3}\,I_4(X) \geq \frac{1}{3}\,I(X)^2.
\en

\vskip2mm
{\bf Proof of Theorem 1.3.} First suppose that the conditions of Proposition 11.2
are fulfilled. Then, by (11.5),
$$
I^{(2)}(X) \geq \frac{1}{3}\int_0^1 L'(t)^4\, dt \geq 
\frac{1}{3}\,\Big(\int_0^1 L'(t)^2\, dt\Big)^2 = \frac{1}{3}\,I(X)^2.
$$
Recalling the representation (11.3) for the functionals $I_p$, (12.1)
follows.

For the finiteness of the integral (11.5), we need to assume that
$I_4(X)$ is finite together with integrability of the function
$(L L'')^2$. In order to give a sufficient condition for this property
to hold, write
$$
L''(t) \, = \,
\frac{d}{dt}\,\frac{f'(F^{-1}(t))}{f(F^{-1}(t))} \, = \,
\frac{f''(F^{-1}(t))}{f(F^{-1}(t))^2} - \frac{f'(F^{-1}(t))^2}{f(F^{-1}(t))^3}
$$
and
\be
L(t) L''(t) \, = \, \frac{f''(F^{-1}(t))}{f(F^{-1}(t))} - 
\Big(\frac{f'(F^{-1}(t))}{f(F^{-1}(t))}\Big)^2.
\en
Using $(x+y)^2 \leq 2x^2 + 2y^2$ ($x,y \in \R$), this implies
$$
L(t)^2 L''(t)^2 \, \leq \, 2\,\Big(\frac{f''(F^{-1}(t))}{f(F^{-1}(t))}\Big)^2 + 2\,
\Big(\frac{f'(F^{-1}(t))}{f(F^{-1}(t))}\Big)^4
$$
and
\be
\int_0^1 L(t)^2 L''(t)^2\,dt \, \leq \, 2\,I^{(2)}(X) + 2\,I_4(X).
\en
Thus, (12.1) is proved provided that the random variable $X$ has 
a density $f$ of class $\mathfrak C^2$ which is positive and is
supported on some interval $(a,b)$ and such that 
$I^{(2)}(X)$ and $I_4(X)$ are finite.

In the general case, the previous step can be applied
to the random variables $X_\ep = X + \ep Z$, $\ep > 0$, assuming that
$Z \sim N(0,1)$ is independent of $X$. In this case, all $X_\ep$ have
positive $C^\infty$-smooth densities with finite $I^{(2)}(X_\ep)$ and 
$I_4(X_\ep)$, according to Corollary 10.2 and Remark 10.3. Hence we get
$$
I^{(2)}(X_\ep) \geq \frac{1}{3}\,I_4(X_\ep) \geq \frac{1}{3}\,I(X_\ep)^2.
$$
Letting here $\ep \rightarrow 0$ and applying (10.2)-(10.3), we arrive at
(12.1).
\qed

\vskip5mm
{\bf Remark 12.1.} We can now explain the last assertion in Proposition 11.2
about the convergence of the integral in (11.5). Assuming that the Fisher-type
information $I^{(2)}(X)$ is finite and applying (12.1), we conclude that the moment 
$I_4(X)$ is finite and hence the integral in (12.3) is finite as well. Thus,
the integral in (11.5) is finite. Conversely, assuming that this integral is finite,
from (12.2) we obtain that
$$
\Big(\frac{f''(F^{-1}(t))}{f(F^{-1}(t))}\Big)^2 \, \leq \, 2\,L(t)^2 L''(t)^2 +
2\,\Big(\frac{f'(F^{-1}(t))}{f(F^{-1}(t))}\Big)^4.
$$
After the integration of this inequality over $0<t<1$, we are led to the
desired conclusion
$$
I^{(2)}(X) \, \leq \, 2 \int_0^1 L(t)^2 L''(t)^2\,dt + 2\,I_4(X) \, < \, \infty.
$$

\vskip5mm
\section{{\bf Stam-type inequality in the case $p \geq 2$}}
\setcounter{equation}{0}

\vskip2mm
\noindent
Recall that the inequality (1.8) of Theorem 1.5 states that,
for all $k = 1,\dots,p-1$, $p \geq 2$,
\be
\frac{1}{I^{(p)}(X+Y)} \geq \frac{1}{I^{(p)}(X)} + \frac{1}{I^{(p)}(Y)} + 
\frac{1}{I^{(k)}(X) I^{(p-k)}(Y)}
\en
whenever the random variables $X$ and $Y$ are independent.
In the case $p=2$, this relation is reduced to
\be
\frac{1}{I^{(2)}(X+Y)} \geq \frac{1}{I^{(2)}(X)} + \frac{1}{I^{(2)}(Y)} + 
\frac{1}{I(X) I(Y)}.
\en
Let us test it on the normal distributions, that is, for
$X \sim N(a_1,\sigma_1^2)$ and $Y \sim N(a_2,\sigma_2^2)$ with 
$a_1,a_2 \in \R$, $\sigma_1,\sigma_2 > 0$. Then
$X+Y \sim N(a_1+a_2,\sigma_1^2 + \sigma_2^2)$, so that
according to Example 2.1,
\bee
I(X) = \frac{1}{\sigma_1^2}, 
 & &
I(Y) = \frac{1}{\sigma_2^2}, \\
I^{(2)}(X) = \frac{2}{\sigma_1^4},
 & &
I^{(2)}(Y) = \frac{2}{\sigma_2^4}, \quad
I^{(2)}(X+Y) = \frac{2}{(\sigma_1^2 + \sigma_2^2)^2}.
\ene
In this case, (13.2) becomes the equality
$$
\frac{(\sigma_1^2 + \sigma_2^2)^2}{2} =
\frac{\sigma_1^4}{2} + \frac{\sigma_2^4}{2} + \sigma_1^2 \sigma_2^2.
$$

\vskip2mm
{\bf Proof of Theorem 1.5.}
One may assume that the random variables $X$ and $Y$ have $C^\infty$-smooth
positive densities $f$ and $g$ with finite Fisher information of all orders up 
to $p$. Indeed, if (13.1) is established under these conditions, in 
the general case one may apply this relation to the random variables
$$
X_\ep = X + \ep Z_1, \quad Y_\ep = X + \ep Z_2 \quad (\ep>0),
$$
assuming that $Z_1$ and $Z_2$ are independent and have a standard
normal distribution. Then $X_\ep + Y_\ep = (X + Y) + \ep \sqrt{2} Z$ with
$Z \sim N(0,1)$, and (13.1) yields
$$
\frac{1}{I^{(p)}(X + Y + \ep \sqrt{2} Z)} \geq 
\frac{1}{I^{(p)}(X_\ep)} + \frac{1}{I^{(p)}(Y_\ep)} + 
\frac{1}{I^{(k)}(X_\ep) I^{(p-k)}(Y_\ep)}.
$$
Letting $\ep \rightarrow \infty$ and applying the continuity property, 
we arrive at the desired relation (13.1) in full generality.

Under the above assumptions, the density of the sum $X+Y$ represents 
the convolution
$$
h(x) = \int_{-\infty}^\infty f(x-y) g(y)\,dy = 
\int_{-\infty}^\infty f(y) g(x-y)\,dy.
$$
By Proposition 4.1, all derivatives of $f$ and $g$ are integrable up 
to order $p$ and are vanishing at infinity up to order $p-1$.
Hence the function $h$ is smooth, everywhere positive, and we have
similar representations for its derivatives of any order $k \leq p$
$$
h^{(k)}(x) = \int_{-\infty}^\infty f^{(k)}(x-y) g(y)\,dy =
\int_{-\infty}^\infty f^{(k)}(y) g(x-y)\,dy.
$$
Differentiating the last integral $p-k$ times, we obtain the $p$-th order derivative
$$
h^{(p)}(x) = \int_{-\infty}^\infty f^{(k)}(x-y) g^{(p-k)}(y)\,dy.
$$

Hence, given real numbers $\alpha_i \geq 0$ such that
$\alpha_0 + \alpha_1 + \dots + \alpha_p = 1$, we have
$$
h^{(p)}(x) = \int_{-\infty}^\infty \, \sum_{k=0}^p 
\alpha_k f^{(k)}(x-y) g^{(p-k)}(y)\,dy.
$$

Introduce the probability measures
$$
\frac{d\mu_x(y)}{dy} = \frac{f(x-y) g(y)}{h(x)}, \quad x \in \R,
$$
and rewrite the above as
$$
\frac{h^{(p)}(x)}{h(x)} = \int_{-\infty}^\infty \, \sum_{k=0}^p 
\alpha_k\,\frac{f^{(k)}(x-y)\, g^{(p-k)}(y)}{f(x-y)\, g(y)}\,d\mu_x(y).
$$
One may now apply Jensen’s inequality, which gives
$$
\Big(\frac{h^{(p)}(x)}{h(x)}\Big)^2 \leq
\int_{-\infty}^\infty \bigg(\sum_{k=0}^p 
\alpha_k\,\frac{f^{(k)}(x-y)\, g^{(p-k)}(y)}{f(x-y)\, g(y)}\bigg)^2\,d\mu_x(y),
$$
or equivalently
\begin{eqnarray}
\frac{h^{(p)}(x)^2}{h(x)} 
 & \leq &
\int_{-\infty}^\infty \bigg(\sum_{k=0}^p \alpha_k\,
\frac{f^{(k)}(x-y)\, g^{(p-k)}(y)}{f(x-y)\, g(y)}\bigg)^2\,f(x-y) g(y)\,dy
 \nonumber \\
 & = &
\sum_{k=0}^p \alpha_k^2 \int_{-\infty}^\infty 
\frac{f^{(k)}(x-y)^2\, g^{(p-k)}(y)^2}{f(x-y)\, g(y)}\,dy \nonumber \\
 & & \hskip-5mm + \ 
\sum_{k \neq l} \alpha_k \alpha_l \int_{-\infty}^\infty 
\frac{f^{(k)}(x-y) f^{(l)}(x-y)}{f(x-y)}\,
\frac{g^{(p-k)}(y) g^{(p-l)}(y)}{g(y)}\, dy.
\end{eqnarray}
Integrating over $x$, we arrive at
\be
I^{(p)}(h) \leq
\sum_{k=0}^p \alpha_k^2\, I^{(k)}(f) I^{(p-k)}(g)  +
\sum_{k \neq l} \alpha_k \alpha_l\, V_{k,l}(f)\,V_{p-k,p-l}(g),
\en
where we use the notation
\be
V_{k,l}(f) = \int_{-\infty}^\infty \frac{f^{(k)}(x) f^{(l)}(x)}{f(x)}\,dx.
\en
Note that these integrals exist and are finite, since,
by Cauchy's inequality,
$$
\int_{-\infty}^\infty \frac{|f^{(k)}(x) f^{(l)}(x)|}{f(x)}\,dx \leq 
\sqrt{I^{(k)}(X) I^{(l)}(X)} < \infty,
$$
and similarly for $g$. This also justifies the integration with respect
to $x$ in (13.3).

If $k=0$ or $l=0$, then the integral in (13.5) is vanishing. Indeed,
in the case $l=0$ and $1 \leq k \leq p$, 
\bee
V_{k,0}(f) 
 & = &
\int_{-\infty}^\infty f^{(k)}(x)\,dx \, = \,
\lim_{T \rightarrow \infty} \int_{-T}^T f^{(k)}(x)\,dx \\
 & = &
\lim_{T \rightarrow \infty} \Big(f^{(k-1)}(T) - f^{(k-1)}(-T)\Big) \, = \, 0,
\ene
where we applied Proposition 4.1. A similar conclusion applies to $g$, and
as a consequence,
\be
V_{k,l}(f)\,V_{p-k,p-l}(g) = 0, \quad {\rm if} \
k = 0, \ k = p, \ l = 0, \ l = p \ \ (k \neq l).
\en

For the setting of Theorem 1.5, we need to restrict ourselves 
to the case where $\alpha_j = 0$ whenever $j \neq 0,k,p$ for a fixed 
$k = 1,\dots,p-1$. Then the second sum in (13.4) contains only the terms, 
which are vanishing, by (13.6). Hence (13.4) is simplified to
$$
I^{(p)}(h) \leq \alpha_0^2\, I^{(p)}(f) + \alpha_p^2\, I^{(p)}(g) + 
\alpha_k^2\, I^{(k)}(f) I^{(p-k)}(g).
$$
Minimizing the right-hand side over all admissible 
$\alpha_i$ yields (13.1).
\qed

\vskip7mm
\section{{\bf Stam-type inequality with Gaussian components}}
\setcounter{equation}{0}

\vskip2mm
\noindent
As we have already mentioned, Theorem 1.5 can be refined in the form of the relation
\be
\frac{1}{I^{(p)}(X+Y)} \, \geq \, \sum_{k=0}^p
\frac{1}{I^{(k)}(X) I^{(p-k)}(Y)},
\en
where one of the independent summands has a normal distribution.
This is a consequence of a more general assertion which 
we state as a lemma.

\vskip5mm
{\bf Lemma 14.1.} {\sl Let $X$ and $Y$ be independent random variables.
Suppose that $X$ has a finite Fisher information $I^{(p)}(X)$ with
a density $f \in \mathfrak C^p$ such that
\be
V_{k,l}(f) = \int_{-\infty}^\infty \frac{f^{(k)}(x) f^{(l)}(x)}{f(x)}\,dx = 0
\quad {\sl for \ all} \ k \neq l \ \ (1 \leq k,l \leq p-1).
\en
Then the relation $(14.1)$ holds true.
}

\vskip5mm
{\bf Proof.} As in the proof of Theorem 1.5, we may assume that
$Y$ has a $C^\infty$-smooth positive density $g$ with finite Fisher 
information of all orders up to $p$, and that the same is true for $X$.
The density $h$ of the sum $X+Y$ has been already shown to satisfy 
the relation (13.4), which is simplified under the condition (14.2) to
\be
I^{(p)}(h) \, \leq \, \sum_{k=0}^p \alpha_k^2\, 
I^{(k)}(f) I^{(p-k)}(g), \quad
\alpha_k>0, \ \alpha_0 + \dots + \alpha_p = 1.
\en

It remains to minimize the right-hand side over all admissible
coefficients $\alpha_k$. So, consider the quadratic function of the form
$$
Q(\alpha_1, \dots, \alpha_p) = 
A_0 \alpha_0^2 + A_1 \alpha_1^2 + \dots + A_p \alpha_p^2, \quad
\alpha_0 = 1 - \alpha_1 - \dots - \alpha_p,
$$
with parameters $A_k > 0$. Its partial derivatives
$\partial_{\alpha_k} Q = -2A_0 \alpha_0 + 2 A_k \alpha_k$
are vanishing if and only if 
$\alpha_k = \frac{A_0}{A_k}\,\alpha_0$, $k = 1,\dots,p$.
Thus, at the point of minimum necessarily 
$$
\alpha_0 \Big(1 + A_0 \sum_{k=1}^p \frac{1}{A_k}\Big) = 1, \quad 
{\rm that \ is}, \quad
\alpha_0 = \frac{1}{A_0}\,\Big(\sum_{k=0}^p \frac{1}{A_k}\Big)^{-1}.
$$
From this we find that
$$
\alpha_k = \frac{1}{A_k}\,
\Big(\sum_{k=0}^s \frac{1}{A_k}\Big)^{-1}, \quad k = 0,1,\dots,p,
$$
and
$$
Q(\alpha_1, \dots, \alpha_p) = \sum_{k=0}^p \frac{1}{A_k}\,
\Big(\sum_{k=0}^p \frac{1}{A_k}\Big)^{-2} = 
\Big(\sum_{k=0}^p \frac{1}{A_k}\Big)^{-1}.
$$
Equivalently, for all $(\alpha_0,\dots,\alpha_p) \in \R^{p+1}$ such that
$\alpha_0 + \dots + \alpha_p = 1$,
$$
Q(\alpha_1,\dots,\alpha_p)^{-1} \geq \sum_{k=0}^p A_k^{-1}.
$$
Hence, (14.1) follows by applying the above inequality with
$A_k = I^{(k)}(X) I^{(p-k)}(Y)$.
\qed

\vskip5mm
{\bf Proof of Theorem 1.6.}
The inequality (14.1) is invariant under all affine transforms
$(X,Y) \rightarrow (c_1,c_2) + \lambda (X,Y)$. 
Hence, when verifying (14.2) in the Gaussian case,
it is sufficient to consider $X$ having a standard normal
distribution with density $\varphi$. Since
$\varphi^{(k)}(x) = (-1)^k H_k(x)\varphi(x)$, the condition (14.2)
is equivalent to the orthogonality of the Chebyshev-Hermite polynomials
in the Hilbert space $L^2(\R,\varphi(x)\,dx)$.
\qed

\vskip7mm
\section{{\bf The Gamma distributions}}
\setcounter{equation}{0}

\vskip2mm
\noindent
Let the random variables $X_n$ have the Gamma 
distributions with $n$ degrees of freedom, that is, with densities
$$
f_n(x) = \frac{x^{n-1}}{\Gamma(n)}\,e^{-x}, \quad x > 0.
$$
As was noticed, $I^{(p)}(X_n)$ is finite if and only if $n>2p$. 
Let us derive the identities (2.1)-(2.2) and then determine the sign for the value 
\be
V_{1,2}(f_n) = \int_0^\infty \frac{f_n'(x) f_n''(x)}{f_n(x)}\,dx.
\en

For the computation of the Fisher-type information, we first note that,
if $u(x) = P(x)\, e^{-x}$ for a smooth function $P$, then
$$
u' = (P' - P)\, e^{-x}, \quad
u'' = (P'' - 2 P' + P)\, e^{-x}, 
$$
so that
\bee
u'^2
 & = &
(P'^2 + P^2 - 2 P' P)\, e^{-2x}, \\
u''^2
 & = &
(P''^2 + 4 P'^2 - 4 P'' P' + P^2 + 2 P'' P - 4 P' P)\, e^{-2x}.
\ene
From this, choosing $P(x) = x^{n-1}$, we have
\bee
\frac{u'^2}{u}
 & = &
\Big(\frac{P'^2}{P} + P - 2P'\Big)\,e^{-x} \, = \,
\big((n-1)^2\,x^{n-3} + x^{n-1} - 2(n-1)\, x^{n-2}\big)\,e^{-x}
\ene
and
$$
\int_0^\infty \frac{u'^2}{u}\,dx = (n-1)^2\,\Gamma(n-2) - \Gamma(n) =
\Gamma(n)\,\Big(\frac{n-1}{n-2} - 1\Big) = \Gamma(n)\,\frac{1}{n-2}.
$$
Hence
$$
I(X_n) = \frac{1}{n-2}, \quad n \geq 2.
$$

Similarly,
\bee
\frac{u''^2}{u}
 & = &
\Big(\frac{P''^2}{P} + 4\,\frac{P'^2}{P} - 4\,\frac{P'' P'}{P} + P +
2 P'' - 4 P'\Big)\,e^{-x} \\
 & = &
\Big((n-1)^2 (n-2)^2\,x^{n-5} + 4 (n-1)^2\, x^{n-3} - 4\,(n-1)^2 (n-2)\,x^{n-4} \\
 & & + \
x^{n-1} + 2(n-1)(n-2)\,x^{n-3} - 4 (n-1)\, x^{n-2}\Big)\,e^{-x}
\ene
and
\bee
\int_0^\infty \frac{u''^2}{u}\,dx 
 & = &
(n-1)^2 (n-2)^2\,\Gamma(n-4) + 4 (n-1)^2\, \Gamma(n-2) + \Gamma(n) \\
 & & \hskip-5mm - \ 
4\,(n-1)^2\,(n-2)\,\Gamma(n-3) + 2(n-1)(n-2)\,\Gamma(n-2) - 4(n-1) \Gamma(n-1) \\
 & = &
\Gamma(n)\,\bigg(\frac{(n-1)(n-2)}{(n-3)(n-4)} - 1 + 4\,\frac{n-1}{n-2} -
4\,\frac{n-1}{n-3}\bigg) \\
 & = &
\Gamma(n)\,\bigg(\frac{(n-1)(n-2)}{(n-3)(n-4)} - 1 - 
\frac{4(n-1)}{(n-2)(n-3)}\bigg).
\ene
After simplifications, we arrive at the desired formula
$$
I^{(2)}(X_n) = \frac{2(n+2)}{(n-2)(n-3)(n-4)}, \quad n \geq 4.
$$

Finally, 
\bee
\frac{u' u''}{u} 
 & = &
\Big(\frac{P'' P'}{P} - 2\,\frac{P'^2}{P} - P'' + 3 P' - P\Big)\,e^{-x} \\
 & = &
\Big((n-1)^2(n-2)\,x^{n-4} - (n-1)(3n-4)\,x^{n-3} + 3(n-1)\,x^{n-2} - x^{n-1}\Big)\,e^{-x},
\ene
so
\bee
\int_0^\infty \frac{u' u''}{u}\,dx 
 & = &
(n-1)^2(n-2)\,\Gamma(n-3) - (n-1)(3n-4)\, \Gamma(n-2) + 2\,\Gamma(n) \\
 & = &
\Gamma(n)\,\Big(\frac{n-1}{n-3} - \frac{3n-4}{n-2} + 2\Big) = 
\Gamma(n)\,\frac{2}{(n-2)(n-3)}.
\ene
Recalling (15.1), it follows that
$$
V_{1,2}(f_n) = \frac{2}{(n-2)(n-3)} > 0, \quad n>3.
$$
This shows that the condition (1.10) is not fulfilled for $f=g=f_n$ with $p=3$.

\vskip5mm
{\bf Acknowledgement.} We would like to thank the referees for careful
reading and useful comments.
Research was partially supported by the NSF grant DMS-2154001 and by
the Hausdorff Research Institute for Mathematics.

\end{document}